\documentclass[preprint]{aastex}

\def\vol#1  {{{#1}{\rm,}\ }}

\def\aj{{AJ}, }  
\def\apj{{ApJ}, } 
\def\apjs{{ApJS}, } 
\def\pasp{{PASP}, }  
\def\mnras{{MNRAS}, } 
\def\aa{{A\&A}, }     

\begin{document}

\title{Evidence for Cluster Evolution from an Improved Measurement of
the Velocity Dispersion and Morphological Fraction of Cluster
1324+3011 at $z = 0.76$}

\author{Lori M. Lubin} 
\affil{Space Telescope Science Institute\altaffilmark{1}, \\ 3700 San
Martin Drive, Baltimore, MD 21218} 
\email{lml@stsci.edu}

\author{J.B. Oke} 
\affil{Palomar Observatory, California Institute of Technology,
Pasadena, CA 91125, and \\ Dominion Astrophysical Observatory, 5071
W. Saanich Road, Victoria, BC V9E 2E7} 
\email{Bev.Oke@nrc.ca}

\author{Marc Postman} 

\affil{Space Telescope Science Institute\altaffilmark{1}, \\ 3700 San
Martin Drive, Baltimore, MD 21218} 
\email{postman@stsci.edu}

\altaffiltext{1}{Space Telescope Science Institute is operated by the
Association of Universities for Research in Astronomy, Inc., under
contract to the National Aeronautics and Space Administration.}

\begin{abstract} 

We have carried out additional spectroscopic observations in the field
of cluster Cl 1324+3011 at $z = 0.76$. Combined with the spectroscopy
presented in Postman, Lubin \& Oke (2001, AJ, 122, 1125), we now have
spectroscopically confirmed 47 cluster members. With this significant
number of redshifts, we measure accurately the cluster velocity
dispersion to be $1016^{+126}_{-93}$ km s$^{-1}$. The distribution of
velocity offsets is consistent with a Gaussian, indicating no
substantial velocity substructure. As previously noted for other
optically-selected clusters at redshifts of $z \gtrsim 0.5$, a
comparison between the X-ray luminosity ($L_x$) and the velocity
dispersion ($\sigma$) of Cl 1324+3011 implies that this cluster is
underluminous in X-rays by a factor of $\sim 3 - 40$ when compared to
the $L_x - \sigma$ relation for local and moderate-redshift clusters.
We also examine the morphologies of those cluster members which have
available high-angular-resolution imaging with the {\it Hubble Space
Telescope} (HST).  There are 22 spectroscopically-confirmed cluster
members within the HST field-of-view.  Twelve of these are visually
classified as early-type (elliptical or S0) galaxies, implying an
early-type fraction of $0.55^{+0.17}_{-0.14}$ in this cluster. This
fraction is a factor of $\sim 1.5$ lower than that observed in nearby
rich clusters. Confirming previous cluster studies, the results for
cluster Cl 1324+3011, combined with morphological studies of other
massive clusters at redshifts of $0 \le z \lesssim 1$, suggest that
the galaxy population in massive clusters is strongly evolving with
redshift. This evolution implies that early-type galaxies are forming
out of the excess of late-type (spiral, irregular, and peculiar)
galaxies over the $\sim 7$ Gyr timescale.

\end{abstract} 


\clearpage

\section{Introduction}

Clusters of galaxies represent the largest scale of fully collapsed,
virialized structures in the universe and provide a powerful probe of
the evolution of structure formation via dynamical collapse.
Therefore, quantifying the abundance and dynamical state of clusters
is key to understanding the formation of clusters and the evolution of
the galaxies within them. With the advent of sensitive X-ray and
optical satellites, such as {\it Einstein}, ROSAT, Chandra, XMM, and
HST, and large ground-based telescopes, such as the Keck 10-m and the
VLT 8-m, it has been possible to extend cluster studies up to
redshifts of $z \sim 1$.  Optical surveys have revealed a only modest
decline in the comoving volume density of rich clusters out to $z \sim
1$ (Gunn, Hoessel \& Oke 1986; Postman et al.\ 1996; Carlberg et al.\
1997; Bahcall \& Fan 1998; Postman et al.\ 2002).  X-ray observations
indicate that the group/cluster X-ray luminosity function does not
evolve significantly to $z \sim 0.8$ for modest X-ray emitters in the
luminosity range of $2 \times 10^{42}$ to $3 \times
10^{44}~h_{50}^{-2}~{\rm ergs~s^{-1}}$ (Ebeling et al.\ 1997; Nichol
et al.\ 1997; Rosati et al.\ 1998; Jones et al.\ 1998); however, there
appears to be a measurable decrease in the number of very X-ray
luminous ($L_{x} \gtrsim 4 \times 10^{44}~h_{50}^{-2}~{\rm
ergs~s^{-1}}$) clusters out to $z \sim 0.9$ (Edge et al.\ 1990; Henry
et al.\ 1992; Bower et al.\ 1994; Jones et al.\ 1998; Henry 2002;
Lewis et al.\ 2002). With these large-area surveys, it has been
possible to study the global X-ray and optical properties of clusters
and use the observed evolution, or lack thereof, to place strong
constraints on the cosmological world model (e.g., Luppino \& Gioia
1995; Carlberg et al.\ 1997; Bahcall \& Fan 1997; Borgani et al.\
2001).

The galaxy populations within massive clusters, and how their
properties evolve with time, have also been the subject of
considerable study.  In local clusters, early-type (elliptical and S0)
galaxies comprise more than 80\% of the total population.  The galaxy
content of clusters is part of the general morphology--density
relation which describes the strong correlation between galaxy
morphology and environment. As the local density increases, the
fraction of early-type galaxies increases, while the fraction of
late-type galaxies decreases (Dressler et al.\ 1980a,b; Postman \&
Geller 1984). Studies of clusters at $z < 0.6$ have revealed
deviations in this universal relation because the morphology and color
of the cluster members are evolving.  One of the most notable changes
is the progressive bluing of the cluster's galaxy population with
redshift. Butcher \& Oemler (1984) found that the fraction of blue
galaxies in a cluster is a strong function of redshift, increasing
from $<5\%$ in local clusters to $\sim 20\%$ in clusters at $z \sim
0.5$.  HST imaging reveals that many of these blue galaxies are either
normal spirals or have peculiar morphologies, resulting in late-type
fractions which are $\sim 3$ times higher than the average current
epoch cluster (Dressler et al.\ 1994, 1997; Couch et al.\ 1994;
Oemler, Dressler \& Butcher 1997; Fabricant, Franx \& van Dokkum
2000).

Even though these clusters show an increased fraction of blue
galaxies, moderate-redshift clusters still contain a population of
early-type galaxies distinguished by extremely red colors and a tight
color--magnitude (CM) relation (a ``red envelope''). Both the mean
color and the CM relation are consistent with that of present--day
ellipticals (Butcher \& Oemler 1984; Arag\'on-Salamanca et al.\ 1991;
Stanford, Eisenhardt \& Dickinson 1995). At $z \gtrsim 0.4$, the red
envelope moves bluewards with redshift (e.g.\ Arag\'on-Salamanca et
al.\ 1993; Lubin 1996; Stanford et al.\ 1997; Stanford, Eisenhardt \&
Dickinson 1998). This color trend is consistent with passive evolution
of an old stellar population formed largely at $z \gtrsim 2$. The
red envelope is reasonably narrow and uniform from cluster to cluster,
implying that the galaxies formed in an episode of star formation
lasting no longer than a few Gyrs (Bower, Lucey \& Ellis 1992a,b;
Ellis et al.\ 1997; Stanford et al.\ 1998; Bower, Kodama \& Terlevich
1998; van Dokkum et al.\ 2001).
 
To probe the properties of clusters and their galaxy populations at
even earlier stages in their development, we have conducted an
extensive observational program to study nine candidate clusters of
galaxies at $z \gtrsim 0.7$ (Oke, Postman \& Lubin 1998; hereafter
Paper I) with the goal of establishing an evolutionary reference
sample of clusters analogous to the MORPHS survey at $z \sim 0.5$
(Smail et al.\ 1997). The cluster sample was chosen from the
optical/near-IR surveys of Gunn, Hoessel \& Oke (1986) and Postman et
al.\ (1996). For each cluster, we have obtained deep $BVRIK'$
photometry from the Keck 10-m and KPNO 4-m telescopes, over 120
low-resolution spectra from a $R$-band magnitude-limited spectroscopic
survey with Keck, and high-angular-resolution imagery from the Wide
Field Planetary Camera 2 (WFPC2) aboard HST. Five papers based on this
survey have been published so far. They include Postman, Lubin \& Oke
(1998; hereafter, Paper II) and Lubin et al.\ (1998; hereafter, Paper
III) which describe the photometric, spectroscopic, and morphological
properties of Cl 1604+4304 ($z = 0.90$) and Cl 0023+0423 ($z = 0.83$).
In addition, Postman, Lubin \& Oke (2001; hereafter, Paper IV)
describe the cluster dynamical properties and the spectrophotometric
properties of the galaxies in Cl 1324+3011 ($z = 0.76$) and Cl
1604+4321 ($z = 0.92$).

In this paper, we present additional spectroscopic observations of the
massive cluster, Cl 1324+3011 at $z = 0.76$, taken at the Keck 10-m
telescope. The goal of these observations is to obtain more cluster
redshifts in order to measure the dynamics of the cluster and the
properties of the cluster galaxy population more accurately. The new
observations are described in \S 2. We have used these new
spectroscopic data, combined with those described in Paper IV, to
improve the measurement of the cluster velocity dispersion and total
mass (\S 3), to make a more accurate comparion between the X-ray and
optical properties of this cluster (\S 4), and to estimate with
reasonable uncertainties the morphological mix of the cluster galaxy
population. The discussion of our results and their implication for
cluster evolution is presented in \S 6.  Unless otherwise noted, we
adopt $h_{65} \equiv H_o/(65~{\rm km~s^{-1}~Mpc^{-1}}) = 1$, $\Omega_o
= 0.2$, and $\Lambda = 0$.

\section{Spectroscopic Observations}

As in our original program, the spectroscopic observations were
conducted using the Low Resolution Imaging Spectrograph (LRIS; Oke et
al.\ 1995) at the W.M. Keck Observatory.  Additional spectra for
galaxies in an approximately $2' \times 7'$ field centered on Cl
1324+3011 were obtained on January 17-19, 2001 at the Keck I
telescope.  We observed two slitmasks, containing 33 and 35 objects
each, with the 400 g/mm grating which is blazed at 8500\AA.  This
grating provides a dispersion of 1.86\AA\ per pixel, and a spectral
coverage of 3800\AA.  The grating angle was set to 7000\AA, so that it
provided coverage from approximately 5500\AA\ to 9500\AA\ in the first
order.  This grating was chosen because of its high spectral
efficiency at 7000\AA\ where we expect the prominent absorption
features, \ion{Ca}{2} H \& K, for galaxies at $z \approx 0.76$.

Broadband optical $BVRI$ and, in some cases, $K'$ data taken as part
of the original survey were used to measure photometric redshifts of
all galaxies within the field using the empirical approach described
in Brunner \& Lubin (2000). This approach works well for almost all
types of galaxies, specifically normal blue, star-forming galaxies and
red, elliptical-like galaxies which are expected to comprise the
majority of cluster members. We note that this method does not work
well for galaxies with unusual spectra, for example quasars or, more
importantly, very strong E+A galaxies, which may comprise 10--20\% of
the cluster population (see e.g.\ Dressler et al.\ 1999). The Brunner
\& Lubin (2002) technique allows us selectively to maximize completion
or minimize contaminaton by tuning the parameters of the
algorithm. Specifically, we can use the photometric redshift and its
associated error to estimate the probability that a given galaxy is a
member of the cluster. Based on the available spectroscopy of this
field, we have chosen a probability threshold which indicates a
completeness level of 76\% and a contamination rate of 28\%. Those
galaxies above this threshold were weighed most heavily when creating
the two slitmasks.  Due to the practicality of assigning objects to a
slitmask, many galaxies which do not meet this probability criterion
are included.  For these cases, the galaxies are weighed according
their $I$-band (or rest-frame $B$-band) luminosity, allowing us to get
a relatively unbaised sample of these galaxies.

Each mask was observed for $4 \times 2400$ sec.  To calibrate the
multi-slit observations, flat-fielding and wavelength calibration were
performed using internal flat-field and arc lamp exposures which were
taken after the science exposures.  Observations of the Oke (1990)
spectrophotometric standard stars Feige 66 and G191B2B were used to
remove the response function of the chip.

Using the procedures described in Paper I, we were able to measure
redshifts for 68\% of the objects observed. Failure to measure a
redshift is mostly due to insufficient signal-to-noise as some of the
targeted objects were as faint as $R \approx 25$.  In Table 1a, we
provide the key photometric and spectroscopic parameters for the new
objects. These parameters are the same ones as given in Table 2 of
Paper IV for the other objects observed in the Cl 1324+3011 field. For
ease of reference in the following sections, we include in Table 1b
the parameters for the original 32 confirmed members of Cl 1324+3011
determined in Paper IV. In Table 1a, a * indicates a confirmed cluster
member based on the dynamical analysis presented in \S 3.

The parameters listed in Tables 1a \& 1b include the Keck object
identification number, the absolute AB magnitudes given in the four
passbands (AB$_B$, AB$_V$, AB$_R$, AB$_I$), geocentric redshift, and
our measure of the redshift quality $Q$ (see Paper I). Measures of the
star-formation rate (column 17), emission line equivalent widths
(columns 12-14), and Balmer jump strengths (columns 15-16) are also
listed.  Emission line intensities are calculated by measuring the
observed flux of the continuum at the line and converting the
equivalent widths to intensities. Negative and positive values
indicate emission and absorption line equivalent widths, respectively.
The typical errors in the equivalent widths are estimated from the
differences between rest-frame equivalent widths derived from two
independent and comparable spectra of the same object.  The
comparisons imply errors of approximately 5\AA\ for the [\ion{O}{2}]
line and $6-8$\AA\ for the H$\beta$ and [\ion{O}{3}]$\lambda5007$
lines.

We also provide a spectral classification (column 10) which is based
on the line strengths of various metal and Balmer absorption
features. Our classification does not include emission line equivalent
widths because they have large intrinsic scatter; therefore, it
differs from the approach used by Dressler et al.\ (1999) who use the
widths of the [\ion{O}{2}] and H$\delta$ lines as their primary
discriminants.  We opt to use our spectral classification scheme
because the four classes encompasss the level at which we can visually
discriminate.  The strengths of the [\ion{O}{2}] are listed separately
and present distinctly different information from the absorption lines
used in the visual spectral classification.  For a full justification
of our clasification scheme and a comparison with that of Dressler et
al.\ (1999), see \S 5.2 of Paper IV. We also refer the reader to
Papers II and IV for a complete description of all parameters listed
here.

\begin{figure}
\plotone{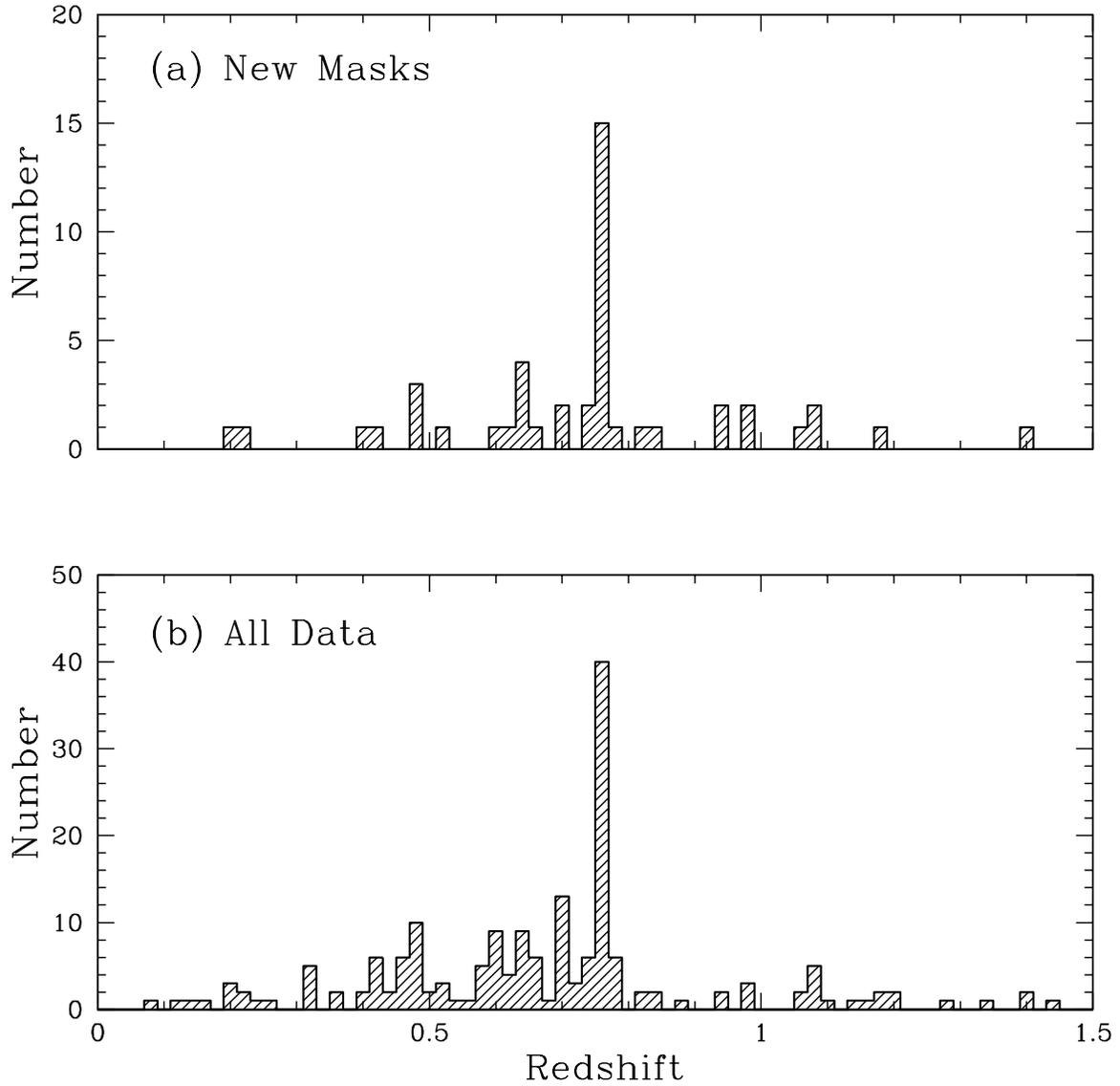}
\caption{(a) Histogram of redshifts (excluding Galactic stars)
obtained from the two new slitmasks covering the Cl 1324+3011
field. (b) Same as (a) but also including the redshift data on this
field available from Paper IV. There is a clear peak in both
histograms at $z \approx 0.76$, the redshift of the cluster.}
\label{zhist}
\end{figure}

In Figure~\ref{zhist}, we show the distribution of redshifts
(excluding Galactic stars) for the galaxies observed in the two new
slitmasks and for all galaxies observed in the Cl 1324+3011 field,
which include the redshifts presented in Paper IV. Both panels show a
clear peak at the measured redshift of the cluster, $z = 0.756$.

\section{Cluster Velocity Dispersion and Mass}

\begin{figure}
\plotone{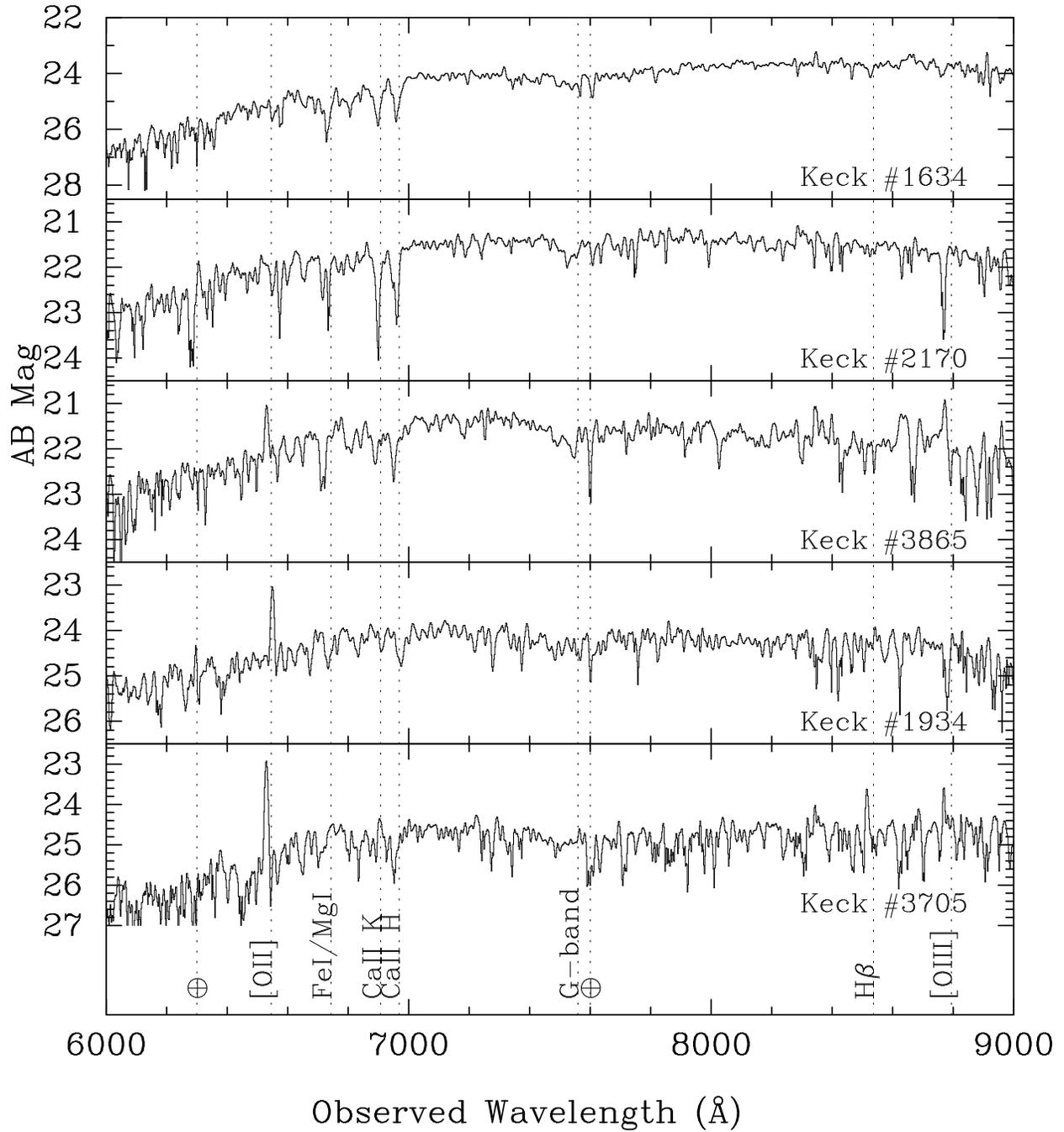}
\caption{LRIS spectra of five of the 15 cluster members determined
from the new spectroscopic observations. The vertical dashed lines
represent the location of of either terrestrial atmospheric lines
($\oplus$) or typical galaxy spectral lines redshifted to the mean
cluster redshift of $\bar{z} = 0.756$. We note that the velocity
dispersion among the cluster members causes significant displacement
of the lines from the mean cluster redshift.}
\label{spec}
\end{figure}

We use the new redshifts, combined with those presented in Paper IV,
to measure a more accurate cluster velocity dispersion. The velocity
dispersion is computed by first defining a broad redshift range,
typically $\Delta z = \pm 0.06$, in which to conduct the calculations.
This range is manually chosen to be centered on the approximate
redshift of the cluster.  We then compute the bi-weight mean and
standard deviation of the velocity distribution (Beers, Flynn, \&
Gebhardt 1990) and identify the galaxy with the largest deviation from
the mean. Velocity offsets from the mean are taken to be $\Delta v =
c(z - \overline{z})/(1 + \overline{z})$ which corrects for
cosmological and relativistic effects. In the case of bi-weight
statistics, $\overline{z}$ is the median of the distribution.  If the
galaxy with the largest velocity deviation differs from the bi-weight
median by either more than 3$\sigma$ or by more than 3500 km s$^{-1}$,
it is excluded, and the computations are redone. The procedure
continues until no further galaxies satisfy the above criteria.  The
3500 km s$^{-1}$ limit is based on extensive data available for low
$z$ clusters (see Papers II and IV for more details). This clipping
procedure is conservative and does not impose a Gaussian distribution
on the final redshift distribution (see e.g.\ CL0023+0423 in Paper
II). Based on this analysis, we find that we have obtained redshifts
for 15 additional cluster members, for a total of 47
spectroscopically-confirmed members.  In Figure~\ref{spec}, we show
spectra of five of the new cluster members. These representative
spectra span the range of spectral properties that are observed in the
cluster members of Cl 1324+3011.

\begin{figure}
\plotone{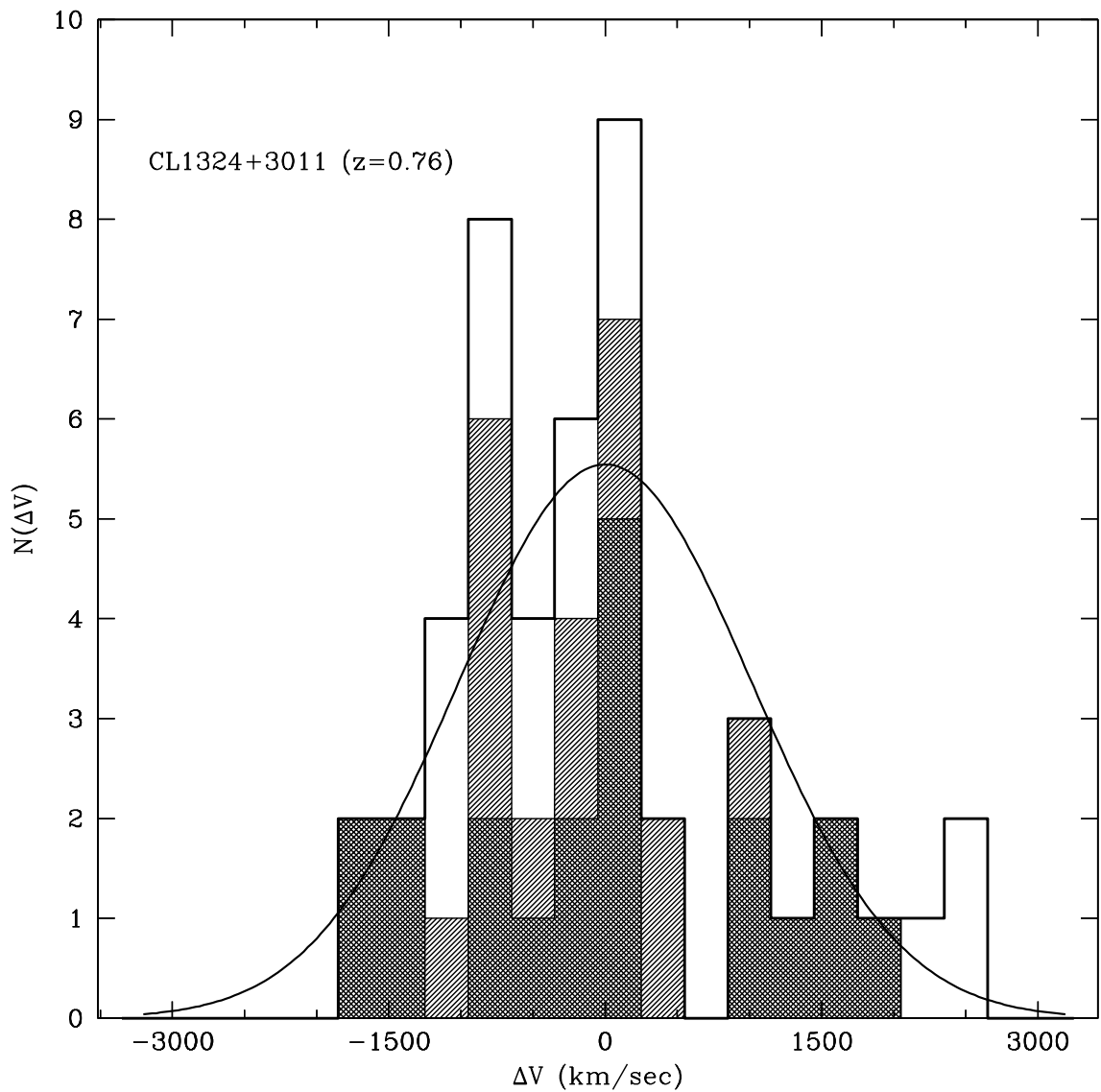}
\caption{Histogram of relativistically-corrected radial velocity
offsets for all confirmed cluster members of Cl 1324+3011. Offsets are
relative to the mean cluster redshift. The best-fit Gaussian
distribution is shown for comparison. The darkest histrogram include
only those galaxies within the central 385 $h_{65}^{-1}$ kpc. The
intermediate shading represents the galaxies within the central 770
$h_{65}^{-1}$ kpc. The unshaded histrogram shows the distribution for
all available data.}
\label{vhist}
\end{figure}

We use the histogram of velocity offsets to measure a new cluster
velocity dispersion of $\sigma = 1016^{+126}_{-93}$ km s$^{-1}$. The
velocity dispersion has been corrected for the redshift measurement
errors which are typically about 200 km s$^{-1}$.  We compute the
uncertainty in the dispersion according to the prescription of Danese,
de Zotta, \& di Tullio (1980), which assumes that the errors in
velocity dispersion can be modeled as a $\chi^2$ distribution and that
a galaxy's velocity deviation from the mean cluster redshift is
independent of the galaxy's mass (i.e., the cluster is virialized). In
Figure~\ref{vhist}, we show the histogram of velocity offsets relative
to the mean cluster redshift for the 47 confirmed cluster members and
the best-fit Gaussian to the distribution.

The new velocity dispersion measurement is completely consistent with
the old measurement, $\sigma = 1058^{+166}_{-114}$ km s$^{-1}$, which
was based on only 32 members (Paper IV). Because of the larger number
of cluster members, the errors in the dispersion have been reduced by
more than 20\%. We use the Kolmogorov-Smirnov (KS) Test to compare the
observed distribution of velocity offsets to a Gaussian model. Based
on the KS statistic, we find that the null hypothesis that the two
distributions are the same can be disproved at only the 72\% level.
Consequently, the distribution of velocity offsets is consistent with
a Gaussian, indicating no obvious or substantial velocity
substructure.  The consistency between the velocity dispersion
measurements even after including $\sim 50\%$ more galaxies suggests
that our measured value is an accurate representation of the true
velocity dispersion of the cluster; however, it is not possible to
determine whether this system is truly virialized based on these data.

\begin{figure}
\plotone{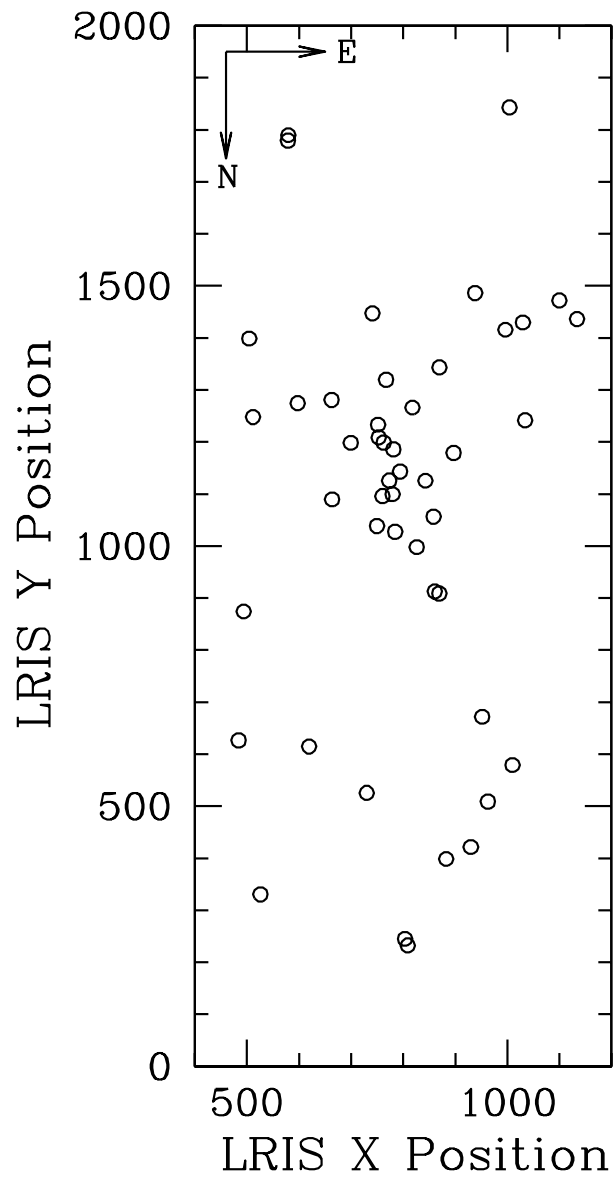}
\caption{Distribution on the sky of the 47 confirmed cluster
members. One LRIS pixel equals 0\farcs{215}. }
\label{xy}
\end{figure}

In Figure~\ref{xy}, we show the distribution on the sky of the
confirmed cluster members. The distribution of cluster members is well
concentrated with 50\% of the cluster members within
$0.45~h_{65}^{-1}$ Mpc of the $I$-band luminosity-weighted cluster
center. The measured harmonic radius is $0.62 \pm 0.01~h_{65}^{-1}$
Mpc, consistent with the wide range of values found in local and
moderate-redshift clusters of galaxies (e.g., Carlberg et al.\ 1996).

As in Paper IV, we use the velocity dispersion and distribution of
cluster members to derive cluster masses based on three virial theorem
mass estimators, the pairwise mass ($M_{PW}$), the projected mass
($M_{PM}$), and the ringwise mass ($M_{RW}$) estimators. The
differences in the mass estimates are due primarily (but not solely)
to the difference in the radius estimators. The pairwise estimator
gives a high weight to close pairs; the other two estimates are less
sensitive to this and, therefore, give more similar values. For the
exact formalisms, see \S 3.1 of Paper II. The derived kinematic
parameters, including mean $z$, dispersion, and mass estimates, are
given in Table 2.  We give these results using all available redshift
data (no radius limit), as well as the results for those galaxies
within the central 385 and 770 $h_{65}^{-1}$ kpc regions. The derived
mass of the cluster, from all of the estimators, is well in excess of
$10^{15}~h_{65}^{-1}~M_{\odot}$, implying that this cluster is typical
of an Abell Richness Class 1 or 2 cluster (e.g.\ Bahcall 1981; Struble
\& Rood 1991).

\section{The $L_x-\sigma$ Relation for High-Redshift Clusters}

In local clusters, which were exclusively identified optically because
of historical reasons (e.g., Abell 1958; Zwicky et al.\ 1961; Abell,
Corwin \& Olowin 1989), the properties of the galaxies and the
intracluster gas are strongly related. In particular, there exist
well-defined correlations between the X-ray properties of the gas,
such as luminosity ($L_x)$ and temperature ($T_x$), and the optical
properties of the galaxies, such as blue luminosity ($L_B$) and
velocity dispersion ($\sigma$). These relations indicate that the
galaxies and gas are in thermal equilibrium, i.e.\ $T_x \propto
\sigma^2$ (e.g.\ Edge \& Stewart 1991).  Moderate-redshift clusters up
to $z \sim 0.5$, selected on the basis of both starlight and X-ray
emission, still exhibit the same X-ray--optical relations (Mushotzky
\& Scharf 1997). However, at higher redshifts of $z \gtrsim 0.5$,
there are indications that at least some massive clusters do not obey
the local relations.  Specifically, optically-selected clusters at $z
\gtrsim 0.5$ (Couch et al.\ 1991; Postman et al.\ 1996; Oke et al.\
1998) do not obey the local $L_{x}-\sigma$ relation (Castander et al.\
1994; Bower et al.\ 1997; Holden et al.\ 1997; Paper IV).  Their X-ray
luminosities are low for their velocity dispersions; thus, they are
underluminous compared to their X-ray--selected counterparts.

\begin{figure}
\plotone{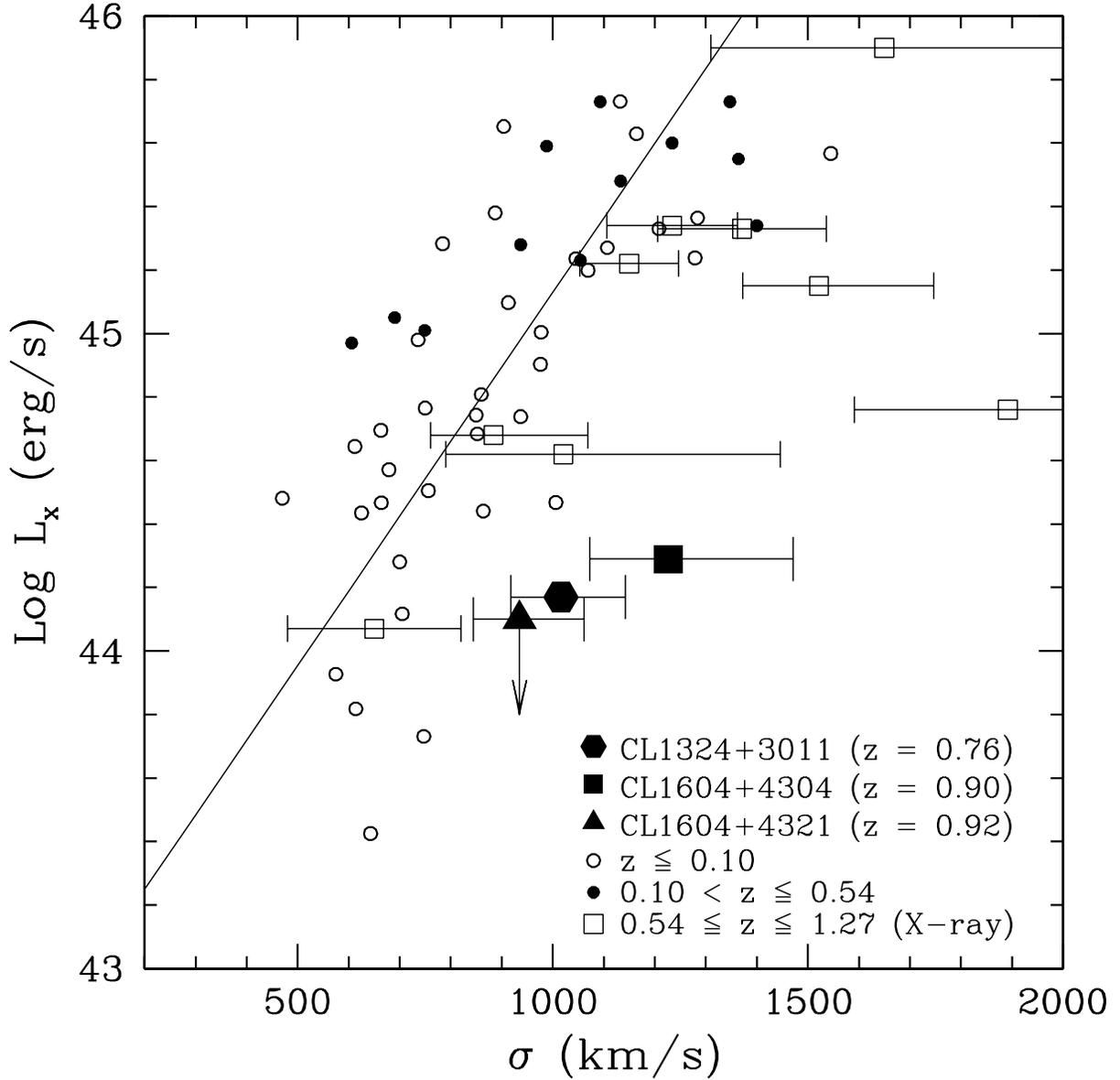}
\caption{Relation between bolometric X-ray luminosity ($h_{50}^{-2}$
ergs s$^{-1}$) and velocity dispersion (km s$^{-1}$). Open and filled
circles indicate clusters at $z \le$ 0.10 and 0.10 $< z \le$ 0.54,
respectively (Mushotzky \& Scharf 1997).  To avoid confusion, we have
not included error bars on these data points; the typical velocity
dispersion error is 86 and 127 km s$^{-1}$ for clusters at $z \le
0.10$ and 0.10 $< z \le$ 0.54, respectively. The filled hexagon
represents the position of Cl 1324+3011 at $z = 0.76$, based on the
new velocity dispersion measurement presented in this paper. The
filled square and filled triangle indicate the two other,
optically--selected clusters, Cl 1604+4304 at $z = 0.90$ and Cl
1604+4321 at $z = 0.92$, presented in Paper IV.  The bolometric X-ray
luminosities of these three clusters are taken from Castander et al.\
(1994). The open squares represent X-ray--selected clusters at 0.54
$\le z \le$ 1.27 (Donahue 1996; Henry et al.\ 1997; Donahue et al.\
1998; Gioia et al.\ 1999; Ebeling et al.\ 2001; Stanford et al.\ 2001,
2002). The solid line indicates the best-fit least-squares relation
for the low and moderate-redshift points.}
\label{lx-sig}
\end{figure}

We can observe this phenomenon in Figure~\ref{lx-sig} which shows the
relation between $L_x$ and $\sigma$ for massive clusters up to $z \sim
1$. The high-redshift ($z \gtrsim 0.5$) sample includes the three
optically-selected clusters from Paper IV (Cl 1324+3011 at $z = 0.76$,
Cl 1604+4304 at $z = 0.90$, and Cl 1604+4321 at $z = 0.91$); however,
we now include Cl 1324+3011 with its improved velocity dispersion
measurement. The X-ray luminosities of these three clusters are taken
from Castander et al.\ (1994). As is obvious from Figure~\ref{lx-sig},
the points representing the three optically--selected clusters at $z
\ge 0.76$ fall at least an order of magnitude below the relation
defined by the low and intermediate-redshift clusters.  For example,
Cl 1324+3011 has a measured X-ray luminosity of $(0.81 \pm 0.25)
\times 10^{44}~h_{50}^{-2}~{\rm ergs~s^{-1}}$ in the $(0.1 - 2.4)$ keV
passband (Castander et al.\ 1994) or a bolometric X-ray luminosity of
$(1.5 \pm 0.5) \times 10^{44}~h_{50}^{-2}~{\rm ergs~s^{-1}}$.  Based
on the $L_x-\sigma$ relation of low-redshift clusters (Mushotzky \&
Scharf 1997) and our improved measurement of the velocity dispersion,
we predict a bolometric X-ray luminosity of $1.5^{+3}_{-1} \times
10^{45}~h_{50}^{-2}~{\rm ergs~s^{-1}}$ (see Paper IV). Consequently,
the X-ray luminosity of Cl 1324+3011 is low by a factor of $\sim 3 -
40$. The results of Cl 1324+3011, and the other optically--selected
clusters, clearly suggest that the relation between the galaxies and
the gas has evolved between redshifts of $z \sim 1$ and the present
day.

In Figure~\ref{lx-sig}, we also show for comparison all high-redshift,
X-ray--selected clusters discovered by {\it Einstein} and ROSAT which
have measured velocity dispersions and X-ray luminosities (Donahue
1996; Henry et al.\ 1997; Donahue et al.\ 1998; Gioia et al.\ 1999;
Ebeling et al.\ 2001; Stanford et al.\ 2001, 2002). Previous X-ray
surveys, such as the {\it Einstein} Medium Sensitivity Survey (Gioia
et al.\ 1990), mainly discovered very X-ray luminous ($L_x \gtrsim 5
\times 10^{44}~h_{50}^{-2}~{\rm ergs~s^{-1}}$) clusters. Consequently,
all but one of the high-redshift, X-ray--selected clusters in
Figure~\ref{lx-sig} are more X-ray luminous than our
optically-selected clusters. We note that deep X-ray surveys, such as
the ROSAT Deep Cluster Survey (Rosati et al.\ 1998) and the Chandra
Deep Field South (Giacconi et al.\ 2001), are now discovering X-ray
underluminous clusters at high redshift. Over a half-dozen clusters at
$z \gtrsim 0.8$ have been identified with very modest X-ray
luminosities of $L_x \sim (0.2 - 4) \times 10^{44}~h_{50}^{-2}~{\rm
ergs~s^{-1}}$ (Henry et al.\ 1997; Giacconi et al.\ 2001; Stanford et
al.\ 2001, 2002; Hashimoto et al.\ 2002; Holden et al.\ 2002).
Unfortunately, we cannot currently compare them to our
optically-selected clusters because the majority of these
X-ray--selected clusters have only a handful of confirmed galaxy
redshifts and, therefore, do not have accurately measured velocity
dispersions.

\section{Cluster Members with HST Imaging}
 
\begin{figure}
\plotone{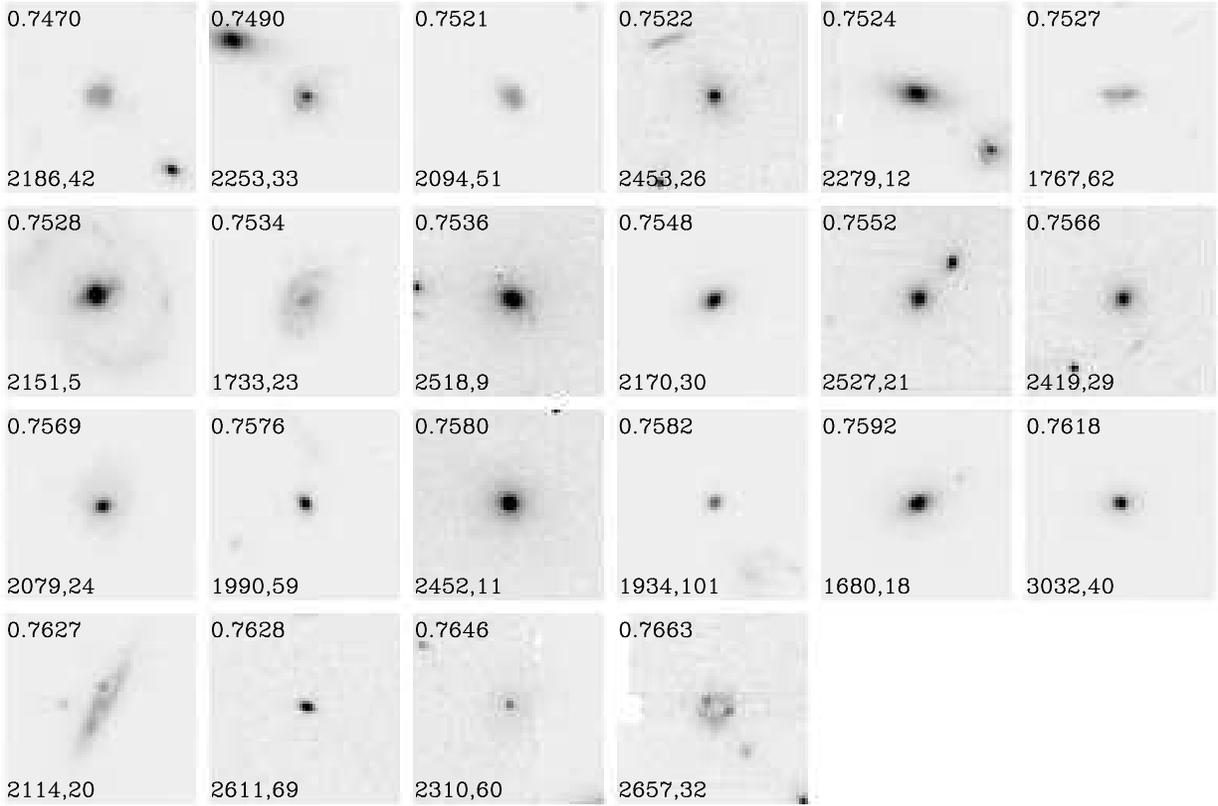}
\caption{All galaxies in the composite F814W image of the Cl 1324+3011
cluster field which are spectroscopically-confirmed members of the
cluster. The galaxies are ordered according to increasing
redshift. The field-of-view of each panel is $5\farcs{98} \times
5\farcs{98}$. The redshift is given in the upper left corner of each
panel. The two numbers at the bottom of each panel indicate the Keck
object identification number (see Paper IV) and the HST identification
number (see Paper V), respectively.}
\label{13hstz}
\end{figure}

The Cl 1324+3011 field was observed by the WFPC2 in the F606W filter
for 16.0 ks and the F814W filter for 32.0 ks in observations taken in
1995 (PI Westphal). We have obtained these data from the archive,
reduced the observations, and measured the properties of all galaxies
within the field. Details on the image reduction, object detection,
and analyses of the galaxy morphological characteristics are described
fully in Lubin et al.\ (2002; hereafter, Paper V). In this paper, we
discuss only those galaxies in the HST image which are
spectroscopically-confirmed cluster members.  From the combined data
of Paper IV and the new redshifts presented here, we find that there
are 22 galaxies in the HST field-of-view which are
spectroscopically-confirmed members of this cluster. In
Figure~\ref{13hstz}, we show postage-stamp images of each of these
galaxies taken from the composite F814W image. We have used data in
this band to make a visual classification of these galaxies based on
the Revised Hubble scheme (e.g., Sandage 1961, Sandage \& Bedke 1994).
The classifications and other relevant information, including some
spectral characteristics, are listed in Table 3 (for more details on
these parameters, see Papers III and V).

\subsection{Spectral Properties}

Twelve of these galaxies are classified as early-type (elliptical or
S0) galaxies. As expected, their spectra are characterized by strong
\ion{Ca}{2} H \& K and G-band absorption features. Consequently, nine
of the twelve galaxies are given a $k$ spectral classification which
is defined as strong \ion{Ca}{2} K, $\lambda3835$, and G band features
with little or no $H_\delta$ absorption (see \S 5.2 of Paper IV). The
remaining three galaxies show more Balmer line activity and are
classified as either $k+a$ or, in one case, $a+k$ (see Table 3). All
but one of these early-type galaxies show little (less than 10\AA\
rest equivalent width) or no [\ion{O}{2}] emission. The
luminosity-normalized star formation rates (SFRNs; see \S 7.2 of Paper
IV) for the early-type galaxies are low, with a mean of 0.09
$M_{\odot}$ yr$^{-1}$ per unit AB$_B$ luminosity. As expected, these
galaxies are also distinguished by their reddest colors, with an
average color of $V-I = 2.61 \pm 0.11$ (see Paper V).  The
optical/near-infrared colors suggest that the stellar populations in
these galaxies formed largely at $z > 2$, consistent with findings for
local and moderate-redshift clusters (Papers II; III; IV; V).

The remaining cluster members are classified as either normal spiral
or irregular/peculiar galaxies. The spectra of these galaxies show
more star-formation activity with moderate-to-strong [\ion{O}{2}],
\ion{O}{3}, and/or $H_{\beta}$ emission (see Table 1 of this paper and
Table 2 of Paper IV). Eight of the ten galaxies have spectral
classification of $k+a$, $a+k$, or $a$ indicating Balmer line activity
(see Table 3). The mean rest-frame equivalent width of the
[\ion{O}{2}] emission is 16\AA, and the mean SFRN is 0.59 $M_{\odot}$
yr$^{-1}$ per unit AB$_B$ luminosity. These galaxies are also
typically bluer than the early-type members, with an average color of
$V-I = 1.67 \pm 0.46$.

\subsection{The Cluster Morphological Fraction}

Based on the 22 spectroscopically-confirmed cluster members with HST
imaging, we measure an early-type fraction in Cl 1324+3011 of
$f_{E+S0} = 0.55^{+0.17}_{-0.14}$. As previously observed in other
high-redshift clusters (e.g., Stanford et al.\ 1997, 1998; van Dokkum
et al.\ 2000; Nelson et al.\ 2001; Paper II; Paper IV), the population
of Cl 1324+3011 still contains a significant number of old, red,
early-type galaxies; however, this fraction is lower by a factor of
$\sim 1.5$, compared to nearby rich clusters (Dressler 1980a,b;
Andreon, Davoust \& Helm 1997).  This decline in early-type fraction
is consistent with other studies of galaxy clusters at redshifts of $z
\approx 0.3 - 0.9$ (Dressler et al.\ 1997; Andreon et al.\ 1997;
Fabricant et al.\ 2000; van Dokkum et al.\ 2000; Paper
III). Consequently, our results, when combined with these previous
observations, imply that the early-type fraction is strongly evolving
with time.

\begin{figure}
\plotone{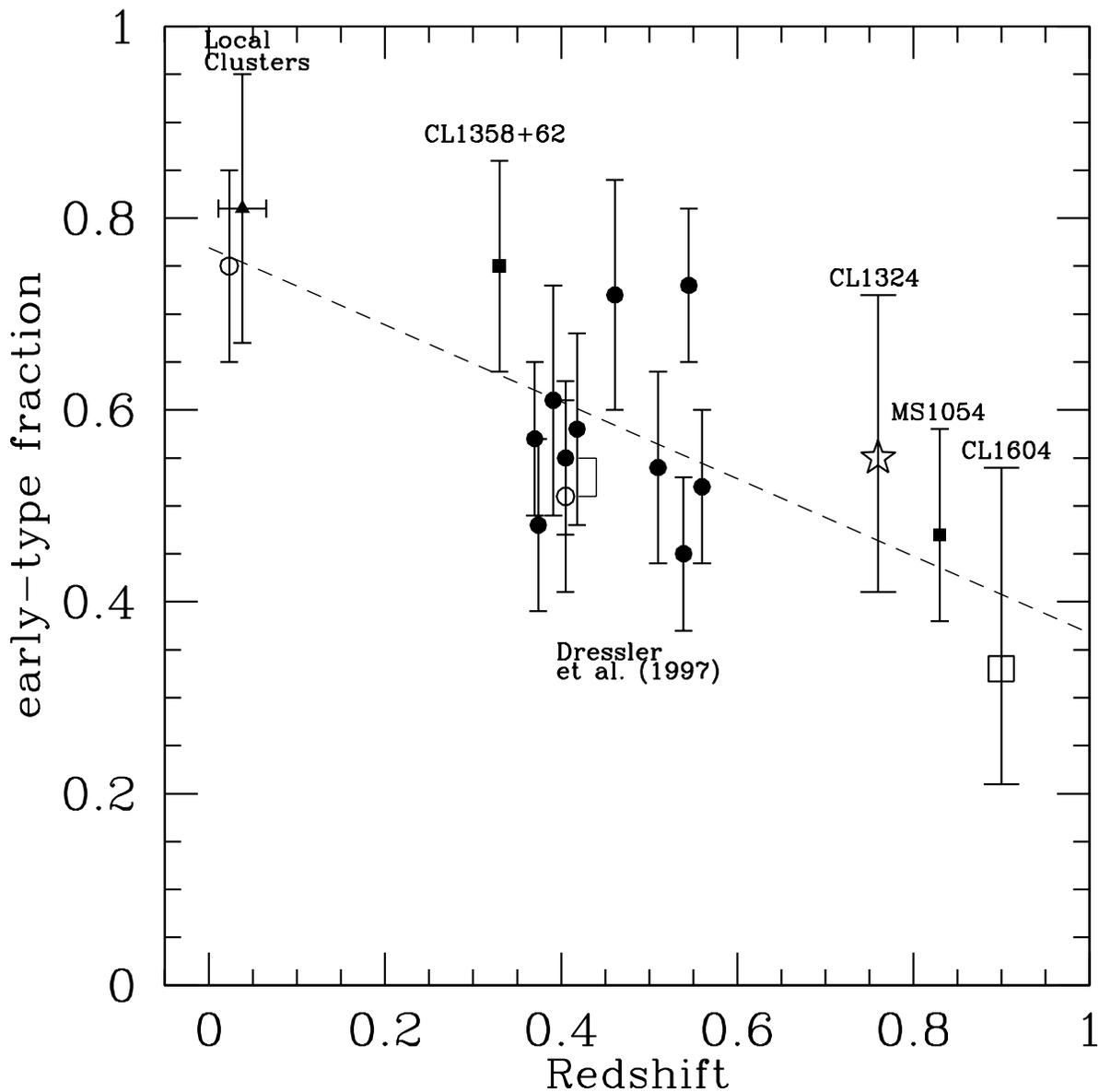}
\caption{Early-type fraction versus cluster redshift. The star
represents the measurement for Cl 1324+3011 ($z = 0.76$) made in this
paper.  Additional data points are taken from Dressler (1980a,b);
Dressler et al.\ (1997); Andreon et al.\ (1997); Fabricant et al.\
(2000); van Dokkum et al.\ (2000); and Paper III. The lines connecting
some data points indicate measurements made of the same cluster by
different authors. Dashed line indicates the best-fit least-squares
line.}
\label{fES0}
\end{figure}

This morphological evolution can clearly be seen in Figure~\ref{fES0}
where we plot the early-type fraction versus cluster redshift for all
currently available morphological studies of galaxy clusters between
$z = 0$ and $z \sim 1$ (Dressler 1980a,b; Dressler et al.\ 1997;
Andreon et al.\ 1997; Fabricant et al.\ 2000; van Dokkum et al.\ 2000;
Paper III), including the new results for Cl 1324+3011 at $z =
0.76$. We note that all of the measurements for the moderate and
high-redshift clusters have been made in roughly the same physical
radius of $r \approx 500~h_{65}^{-1}$ kpc, corresponding to one WPFC2
pointing. Unlike our measurement and the others made at $z > 0.6$,
Dressler et al.\ (1997) do not include galaxies which appeared
extremely chaotic or insufficiently detailed to fit into a Revised
Hubble type in their determination of the early-type fractions in
their sample of ten clusters at $z = 0.37 - 0.56$. These ``peculiar''
galaxies account for only a few percent of the total and, therefore,
do not affect their overall measurements (Dressler et al.\ 1997; Smail
et al.\ 1997).

The local data points are derived from two sources.  First, we use the
data, presented in Dressler (1980a,b) and reanalyzed in Dressler et
al. (1997), of the morphological fraction versus local surface density
for a sample of 55 clusters at $z \sim 0.04$. Based on data given in
Table 2 of Dressler et al.\ (1997), we calculate the median early-type
fraction (and its associated uncertainty) in local clusters within
regions of local surface density [i.e., log $\Sigma$
(galaxies/Mpc$^{2}$) $> 1.3$] which are similar to those probed in the
moderate and high-redshift clusters (see Dressler et al.\ 1997; Paper
V).  Second, we plot the measurement of the morphological mix in the
central region ($r \approx 1~h_{65}^{-1}$ Mpc) of the Coma cluster at
$z = 0.023$ made by Andreon et al.\ (1997). Both studies yield similar
measures of the early-type fraction ($\sim 70-80\%$) in the central
regions of nearby rich clusters.

Figure~\ref{fES0} clearly shows that the early-type fraction declines
with redshift.  At $z \sim 1$, the fraction is lower by a factor of
$\sim 1.5-2.0$, compared to local clusters.  Consequently, the
fraction of late-type (spiral, irregular, and peculiar) galaxies
increases with redshift. This evolution implies that early-type
galaxies are forming out of the excess of late-type galaxies over a
timescale of $\sim 7$ Gyr. The observed evolution in the morphologies
of the cluster galaxies is completely consistent with the photometric
and spectral analyses of massive clusters at redshifts between $z \sim
1$ and the present. In particular, Butcher \& Oemler (1984) were the
first to detect a progressive bluing of the cluster's galaxy
population with redshift.  They found that the fraction of blue
galaxies in a cluster is a strong function of redshift, increasing
from $<5\%$ in nearby clusters to $\sim 20\%$ in clusters at $z \sim
0.5$.  HST imagery revealed that most of these blue galaxies are
either normal spirals or have peculiar morphologies, resulting in
late-type fractions which are $\sim 3$ times higher than the average
current epoch cluster (Dressler et al.\ 1994, 1997; Couch et al.\
1994; Oemler, Dressler \& Butcher 1997; Fabricant et al.\ 2000). In
addition, our spectral analysis of three clusters at $0.76 \le z \le
0.92$ shows a similar trend at even higher redshifts.  We find a much
higher fraction of blue, star-forming galaxies in the cluster
population. Within the central $1.5~h_{65}^{-1}$ Mpc, the fraction of
active ([\ion{O}{2}] rest equivalent widths greater than 15\AA)
galaxies is 45\% (Paper IV). This fraction is substantially higher
than the 10--20\% active galaxy component seen in the centers of
clusters at $0.2 < z < 0.55$ (Balogh et al.\ 1997).

\section{Discussion}

We have obtained additional spectroscopy for galaxies in the field of
the massive cluster Cl 1324+3011 at $z = 0.76$. With these new data
and those previously taken as part of our original cluster survey
(Paper IV), we have now spectroscopically-confirmed 47 cluster
members, of which 22 have available high-angular-resolution HST
imaging and, therefore, have measured morphological properties.  We
have used these data to measure accurately the cluster velocity
dispersion ($\sigma$) and the fraction of early-type galaxies in the
cluster galaxy population ($f_{E+S0}$).  We find $\sigma =
1016^{+126}_{-93}$ km s$^{-1}$ and $f_{E+S0} = 0.55^{+0.17}_{-0.14}$.
Both of these quantities imply significant differences between Cl
1324+3011 and the average properties of local clusters of galaxies.

Firstly, Cl 1324+3011 does not follow the relation between X-ray
luminosity and velocity dispersion observed in nearby and
moderate-redshift clusters. We find that the X-ray luminosity of Cl
1324+3011 is low by a factor of $\sim 3 - 40$ for its velocity
dispersion and, therefore, its estimated total mass. This cluster is
typical of optically-selected clusters at redshifts of $z \gtrsim 0.5$
and seems to imply that the galaxies and the gas are no longer in
thermal equilibrium. We expect such behavior if these clusters are
still dynamically young. For example, at these early epochs, clusters
of galaxies may still be in the process of merging.  This process can
result in a velocity dispersion which is inflated due to infalling
matter or large scale structure, artificially implying a high mass.
However, the X-ray emission can also evolve in a significant way if
(1) subclumps in the process of cluster formation are not yet fully
virialized or (2) subclumps are currently in the process of merging.
In the former case, the gas will reside in lower-temperature (and
lower-density) subclumps, substantially reducing its X-ray emission
relative to the total mass within the region. In the latter case, the
cluster would have a significant excess luminosity and temperature
over that of non-merging clusters, which results from higher-density
shocks between the merging cores (Ricker 1998; Ricker \& Sarazin 2001;
Sarazin 2001). These enhancements can be over 4--10 times in
lumonisity and up to 3 times in temperature, and the effect can last
at least one crossing time ($\sim$ 1--2 Gyr).

There is, in fact, strong observational evidence that a large fraction
(up to 75\%) of clusters at $z \sim 1$ are still in the process of
forming and are, therefore, dynamically young (see e.g.\ Henry 2002).
The signatures of this formation include double peaks, strong
substructure, and/or a filamentary appearance in the distribution of
galaxies, gas, and total mass (e.g., Lubin \& Postman 1996; Henry et
al.\ 1997; Luppino \& Kaiser 1997; Donahue et al.\ 1998; Gioia et al.\
1999; Della Ceca et al.\ 2000; Ebeling et al.\ 2000; Jeltema et al.\
2001; Stanford et al.\ 2001; Hashimoto et al.\ 2002).

Despite evidence suggesting that most clusters at $z > 0.7$ are not
relaxed, there is no evidence that the average relation between X-ray
luminosity and temperature has evolved with redshift. Using a sample
of 11 X-ray--selected clusters, Holden et al.\ (2002) find no
statistically significant evolution in the slope or zero-point of the
$L_x-T_x$ relation at a median redshift of $z = 0.83$ (see also
Donahue et al.\ 1999; Della Ceca et al.\ 2000; Fairley et al.\ 2000;
Arnaud, Aghanim \& Neumann 2002).  However, there may be preliminary
indications that some clusters (for example, the double-peaked cluster
RXJ 0152.7-1357 at $z = 0.833$; Ebeling et al.\ 2000; Della Ceca et
al.\ 2000; Lewis et al.\ 2002) do fall well off the $L_x-T_x$
relation.  This behavior would be expected if the dynamical and
evolutionary state of clusters at these early epochs is diverse, as
indicated by some optical, X-ray, and weak-lensing studies.

As of yet, we have not been able to determine where optically-selected
clusters at high redshift fall on the $L_x-T_x$ relation as none have
been the subject of X-ray spectral observations. However, this
situation will soon be rectified for Cl 1324+3011 as we are currently
in the process of analyzing a 40 ksec XMM observation of this
cluster. For the first time, we will have accurate information on the
luminosity, temperature, and mass of an optically-selected cluster at
$z \sim 0.8$. These data will go far in determining what physical
processes are responsible for the differences between X-ray and
optically-selected clusters at these redshifts.

Secondly, we measure an early-type fraction in Cl 1324+3011 which is
lower by a factor of $\sim 1.5$ compared to that observed in nearby
rich clusters (Dressler 1980a,b; Andreon et al.\ 1997). A decrease in
the number of early-type galaxies was first observed by Dressler et
al.\ (1997), Andreon et al.\ (1997), Fabricant et al.\ (2000), and van
Dokkum et al.\ (2000) in massive clusters with redshifts between $z
\approx 0.3 - 0.8$. The results for Cl 1324+3011, combined with these
previous studies, show a general trend of decreasing early-type
fraction with redshift. The most significant question raised by this
evolution is what physical mechanisms associated with the cluster
environment can transform disk galaxies into spheroids. Over the
years, at least three plausible mechanisms have been suggested. The
first is galaxy harassment which involves a high-speed encounter of a
faint spiral (Sc or Sd) galaxy with the cluster tidal field or a
brighter cluster galaxy. The tidal shock strips the stellar disk,
producing a remnant consistent with the mass profiles of dwarf
ellipticals (Moore, Lake \& Katz 1998).  The second is ram pressure
stripping by the intracluster medium as an early spiral (Sa or Sb)
galaxy enters the cluster potential. This interaction can result in
either a burst or truncation of star formation (Gunn \& Gott 1972;
Abadi, Moore \& Bower 1999; Poggianti et al.\ 1999).  In both
scenarios, the gas is eliminated, and the disk fades. The final
mechanism is galaxy--galaxy merging. The morphological signatures of a
merger between two disk galaxies, such a long, bright tidal tail,
would be visible for only 1--2 Gyrs.  After this time, the tidal
debris becomes indistinguishable from the main merger remnant, a
morphologically--normal elliptical galaxy (Schweizer 1986; Hibbard et
al.\ 1994; Mihos 1995). Merging is unlikely to occur in the cluster
environment because the velocities are too high; however, small groups
of galaxies being accreted into the cluster are ideal sites for
merging because the group velocity dispersion is on the order of the
internal velocity of the member galaxies (Aarseth \& Fall 1980; Barnes
1985; Merritt 1985).

At least two of these processes require infall of individual galaxies
and/or galaxy groups into the cluster environment. Since the field
population is comprised mainly of spiral and irregular galaxies (e.g\
Driver et al.\ 1995; Abraham et al.\ 1996), infall is the natural
means for providing a reservoir of late-type galaxies. Simulations
predict that clusters should form at the intersections of large scale
sheets and filaments (Shandarin \& Zel'dovich 1989) through the
accretion of smaller subclumps from the surrounding environment
(Evrard 1990; Lacey \& Cole 1993, 1994). Semi-analytic models show
that the accretion rates increase strongly with look-back time. In
early stages of cluster formation, the infall rate is up to a factor
of $\sim 4$ higher than in present-day clusters (Kauffmann 1995).
Observationally, the strong substructure and filamentary nature in a
large fraction of clusters at $z > 0.7$ (see above) suggests that
clusters are actively forming at these redshifts.

These results indicate that infall regions are likely to be active
sites of galaxy evolution.  Unfortunately, the field-of-view of WFPC2,
which is only $160'' \times 160''$, has strongly limited morphological
studies in these regions. Because of these constraints, only one
morphological study of a high-redshift cluster has extended beyond the
central $0.5~h_{65}^{-1}$ Mpc (i.e.\ a single WFPC2 pointing), that of
the very X-ray luminous cluster MS 1054.4-0321 at $z = 0.83$ (van
Dokkum et al.\ 1999, 2000). This WFPC2 study, which covered a 2.7
$h_{65}^{-1} \times$ 1.8 $h_{65}^{-1}$ Mpc$^{2}$ area in the F606W and
F814W filters, has revealed some stunning results. The authors find
that 17\% of the $L \ge L_{*}$ cluster population is comprised of
ongoing mergers which will likely evolve into luminous ($\sim 2
L_{*}$) early-type galaxies. These mergers are preferentially found in
the outskirts of the cluster, indicating they probably reside in cold
infalling clumps. If the galaxy population in MS1054.4-0321 is typical
for its redshift, 50\% of the early-type galaxies in present-day
clusters have experienced a major merger at $z < 1$. The crucial
question is whether MS1054.4-0321 is typical of other clusters at
these redshifts.  In order to convincingly discriminate between the
physical processes which can transform galaxy morphologies in this
environment, it is necessary to study galaxy properties at large radii
for a statistical sample of high-redshift clusters. We expect
significant progress to be made on this front with the wide-area
capability of the Advanced Camera for Surveys which was placed aboard
HST this year.

\acknowledgments

We thank the anonymous referee for very useful comments on the
text. We also thank C.D.\ Fassnacht for aid in conducting the
observations used in this paper and for use of his FITS plotting
program.  Observational material for this paper was obtained at the
W. M. Keck Observatory, which is operated as a scientific partnership
between the California Institute of Technology, the University of
California, and the National Aeronautics and Space Administration.  It
was made possible by the generous financial support of the W. M. Keck
Foundation.

\clearpage

\hoffset -0.9in
\begin{deluxetable}{lrrrrrrrrcrrrrrrrr}
\tablewidth{0pt}
\tabletypesize{\scriptsize}
\tablenum{1a}
\tablecaption{Photometric and Spectroscopic Data from New Cl 1324+3011 Observations}
\tablehead{
\colhead{ID\#} &
\colhead{$AB_B$} &
\colhead{$AB_V$} &
\colhead{$AB_R$} &
\colhead{$AB_I$} &
\colhead{z} &
\colhead{Q} &
\colhead{b} &
\colhead{tau} &
\colhead{Sp} &
\colhead{Age $\tau0.6$} &
\colhead{[OII]} &
\colhead{H$\beta$} &
\colhead{[OIII]} &
\colhead{$J_l$} &
\colhead{$J_u$} &
\colhead{SFR} &
\colhead{$M_{ABB}$}\\
\colhead{(1)} &
\colhead{(2)} &
\colhead{(3)} &
\colhead{(4)} &
\colhead{(5)} &
\colhead{(6)} &
\colhead{(7)} &
\colhead{(8)} &
\colhead{(9)} &
\colhead{(10)} &
\colhead{(11)} &
\colhead{(12)} &
\colhead{(13)} &
\colhead{(14)} &
\colhead{(15)} &
\colhead{(16)} &
\colhead{(17)} &
\colhead{(18)}}
\startdata
  37 & 22.73 & 21.84 & 21.49 & 21.11 & 0.2166 &	 4 &   5.49 &  5.0 & \nodata &   2.4 &  \nodata &  20.3 &   14.4  & \nodata  & \nodata  &	 0.000 &   -18.11 \\
 292 & 25.52 & 24.74 & 23.72 & 23.54 & 1.1838 &	 1 &   6.52 &  1.0 & \nodata &   2.6 &  \nodata &  \nodata &   \nodata  & \nodata  & \nodata  &	\nodata &   -21.49 \\
 297 & 24.47 & 23.97 & 23.31 & 22.91 & 0.4852 &	 4 &   5.88 &  5.0 &  $k$    &   1.7 &  \nodata &  -8.8 &  -27.1  & \nodata  & \nodata  &	 0.533 &   -18.45 \\
 469 & 25.53 & 25.12 & 24.13 & 23.38 & 1.0798 &	 2 &   7.73 &  0.6 & \nodata &   3.0 & -82.4 &  \nodata &   \nodata  & \nodata  & \nodata  &	 5.439 &   -20.79 \\
 478 &\nodata& 25.12 & 23.51 & 22.48 & 0.8455 &	 1 &  \nodata & \nodata & $k+a$&   0.1 & -38.6 &  \nodata &   \nodata  &  0.51  & -0.01  &	 0.000 &  \nodata \\
 546 & 23.48 & 22.99 & 22.62 & 22.10 & 0.2033 &	 4 &   4.99 & 10.0 & \nodata &   1.9 &  \nodata & -38.9 & -182.3  & \nodata  & \nodata  &	 0.915 &   -16.93 \\
 570 & 24.90 & 25.57 & 23.10 & 21.99 & 0.3919 &	 2 &  10.01 &  0.2 &  $k$ &   9.0 &  \nodata &   2.6 &   -2.0  & \nodata  & \nodata  &	 0.034 &   -17.98 \\
 652 & 26.01 & 28.74 & 23.70 & 23.05 & 9.0000 &	 0 &  \nodata & \nodata & \nodata &  \nodata &  \nodata &  \nodata &   \nodata  & \nodata  & \nodata  &	\nodata &    \nodata \\
 715* & 27.29 & 26.45 & 23.23 & 21.38 & 0.7509 &	 3 &  15.79 & \nodata & $k+a$&  11.0 &   1.0 &  -5.1 &    1.4  &  1.54  & 	0.58  &	 0.000 &   -20.70 \\
 762 & 26.04 & 25.34 & 23.85 & 22.72 & 9.0000 &	 0 &  10.63 & \nodata & \nodata &  \nodata &  \nodata &  \nodata &   \nodata  & \nodata  & \nodata  &	\nodata &   -17.16 \\
 828 & 29.00 & 27.37 & 21.19 & 19.41 & 0.4219 &	 4 &  14.76 & \nodata & $k$  &  11.0 &  \nodata &   2.5 &    1.6  & \nodata  & \nodata  &	 0.000 &   -20.39 \\
 835 & 26.09 & 25.54 & 24.30 & 23.66 & 1.3996 &	 2 &   8.13 &  0.6 & \nodata &   3.2 & -32.9 &  \nodata &   \nodata  & \nodata  & \nodata  &	 5.542 &    \nodata \\
 934 &\nodata& 26.73 & 22.68 & 21.31 & 0.7849 &	 3 &  \nodata & \nodata & $a+k$&   0.1 & -13.6 &  -2.4 &  -10.2  &  0.91  & 	0.17  &	 0.000 &  \nodata \\
 985 & 24.13 & 24.03 & 23.55 & 22.41 & 1.0530 &	 2 &   6.86 &  1.0 &$a+k$ &   2.6 & -22.0 &  \nodata &   \nodata  & \nodata  & \nodata  &	 2.517 &   -21.32 \\
1040 & 22.10 & 21.79 & 21.37 & 21.09 & 0.5952 &	 4 &   4.11 & long &$a+k$ &   1.1 & -57.3 & -15.1 &  -47.2  &  0.99  & 	0.48  &	 6.275 &   -20.94 \\
1050 & 23.01 & 22.63 & 22.30 & 21.94 & 0.4815 &	 4 &   4.04 & long &$k+a$ &   1.0 & -64.6 & -11.4 &  -34.2  &  0.81  & 	0.43  &	 1.651 &   -19.49 \\
1085 & 24.16 & 23.73 & 23.82 & 23.38 & 0.4810 &	 3 &   2.32 & long & \nodata &   0.3 &  -4.1 &   3.0 &  -22.1  & \nodata  & \nodata  &	 0.033 &   -18.02 \\
1154 & 25.16 & 24.30 & 23.84 & 22.82 & 0.6444 &	 4 &   8.20 &  1.5 &$k$   &   2.6 & -45.2 &   6.8 &  -17.4  &  1.96  & 	0.81  &	 0.651 &   -18.99 \\
1247 & 22.36 & 21.01 & 24.09 & 23.50 & 9.0000 &  0 &   \nodata & \nodata & \nodata &  \nodata &  \nodata &  \nodata &   \nodata  & \nodata  & \nodata  &	\nodata &    \nodata \\
1484 & 25.93 & 24.73 & 25.57 & 23.93 & 9.0000 &	 0 &   5.81 & \nodata & \nodata &  \nodata &  \nodata &  \nodata &   \nodata  & \nodata  & \nodata  &	\nodata & \nodata  \\
1499 & 23.95 & 23.91 & 23.47 & 22.74 & 9.0000 &	 0 &   4.22 & \nodata & \nodata &  \nodata &  \nodata &  \nodata &   \nodata  & \nodata  & \nodata  &	\nodata &    \nodata \\
1593* & 24.94 & 24.48 & 23.95 & 22.96 & 0.7552 &	 4 &   7.40 &  1.5 &$a+k$ &   2.2 & -61.1 &  -2.2 &  -16.6  &  1.02  & 	0.24  &	 1.329 &   -19.43 \\
1634* & 25.59 & 25.68 & 23.34 & 21.60 & 0.7548 &	 4 &  14.62 &  1.0 &$k+a$ &  11.0 &  \nodata &  \nodata &   \nodata  & \nodata  & \nodata  &	\nodata &   -20.58 \\
1713 & 25.31 & 25.09 & 24.82 & 23.68 & 9.0000 &	 0 &   5.92 & \nodata & \nodata &  \nodata &  \nodata &  \nodata &   \nodata  & \nodata  & \nodata  &	\nodata &  \nodata \\
1721* & 24.26 & 23.62 & 22.89 & 21.86 & 0.7536 &	 2 &   8.71 &  1.5 &$a+k$ &   2.8 & -22.3 &  -9.3 &    7.3  &  0.80  & 	0.07  &	 1.261 &   -20.55 \\
1792* & 24.44 & 24.02 & 23.74 & 22.99 & 0.7576 &	 1 &   5.09 &  3.5 &$a$   &   1.6 & -19.3 &  \nodata &   \nodata  &  0.37  & 	0.06  &	 0.508 &   -19.52 \\
1889 & 25.50 & 24.53 & 22.74 & 21.23 & 9.0000 &	 0 &  13.02 & \nodata & \nodata &  \nodata &  \nodata &  \nodata &   \nodata  & \nodata  & \nodata  &	\nodata &   -21.65 \\
1934* & 24.85 & 24.63 & 23.89 & 22.66 & 0.7582 &	 4 &   8.58 &  1.5 &$a+k$ &   2.6 & -21.1 &  -3.2 &   -1.6  &  0.99  & 	0.16  &	 0.525 &   -19.64 \\
1978 & 22.34 & 21.98 & 21.77 & 21.00 & 0.9418 &	 4 &   4.78 &  2.0 &$a+k$ &   1.7 & -10.4 &  \nodata &   \nodata  & \nodata  & \nodata  &	 3.474 &   -22.33 \\
2079* & 23.88 & 23.25 & 22.58 & 21.38 & 0.7569 &	 4 &   9.54 &  1.0 &$k$   &   3.0 &  -6.7 &  -0.5 &   \nodata  &  0.85  & 	0.33  &	 0.541 &   -20.95 \\
2094* & 24.27 & 23.89 & 23.30 & 22.57 & 0.7521 &	 4 &   6.26 &  2.0 &$a$   &   2.0 & -21.4 &  -6.3 &   \nodata  &  0.97  & 	0.30  &	 0.800 &   -19.98 \\
2170* & 26.90 & 23.92 & 23.07 & 21.46 & 0.7548 &	 4 &  13.36 &  0.2 &$k$   &   7.0 &   0.8 &   0.4 &    4.4  &  1.03  & 	0.49  &	 0.000 &   -20.76 \\
2189 & 23.34 & 23.00 & 22.72 & 22.05 & 1.0783 &	 2 &   4.50 &  1.2 &$a$   &   2.2 & -20.4 &  \nodata &   \nodata  & \nodata  & \nodata  &	 4.579 &   -22.00 \\
2280* & 25.23 & 24.97 & 24.47 & 23.10 & 0.7471 &	 4 &   8.38 &  1.5 &$k$   &   2.6 &  \nodata &  \nodata &   \nodata  & \nodata  & \nodata  &	\nodata &   -19.09 \\
2445 & 24.15 & 23.72 & 23.72 & 22.93 & 0.6377 &	 4 &   3.95 & long &$a+k$ &   1.1 & -62.9 &  -1.3 &  -22.6  &  1.20  & 	0.41  &	 1.151 &   -18.98 \\
2518* & 23.99 & 22.22 & 21.33 & 19.81 & 0.7536 &	 4 &  12.69 &  0.6 &$k$   &   5.0 &   2.4 &   2.9 &   -7.0  &  0.85  & 	0.46  &	 0.000 &   -22.42 \\
2542* & 27.03 & 24.61 & 23.94 & 22.79 & 0.7671 &	 3 &  10.10 &  1.0 &$a$   &   3.5 &  -8.0 &  -2.4 &   \nodata  &  1.41  & 	0.44  &	 0.186 &   -19.68 \\
2709 & 22.75 & 22.19 & 21.60 & 20.65 & 0.6514 &	 4 &   7.77 &  2.0 &$a$   &   2.4 &   4.4 &   2.1 &   -0.6  & \nodata  & 	0.57  &	 0.000 &   -21.23 \\
2711 & 24.12 & 23.26 & 22.68 & 21.12 & 9.0000 &	 0 &  11.98 & \nodata & \nodata &  \nodata &  \nodata &  \nodata &   \nodata  & \nodata  & \nodata  &	\nodata & \nodata \\
2846 & 25.82 & 24.63 & 23.81 & 22.24 & 0.6200 &	 2 &  12.96 &  0.2 &$a$   &   9.0 & -55.8 &   5.6 &  -12.8  &  1.32  & 	0.54  &	 0.678 &   -19.13 \\
2855 & 24.35 & 23.66 & 23.15 & 22.44 & 0.9376 &	 2 &   6.64 &  1.5 & \nodata &   2.4 & -13.6 &  \nodata &   \nodata  &  0.01  & -0.00  &	 1.275 &   -21.07 \\
2951 & 24.65 & 24.67 & 24.33 & 23.43 & 0.6922 &	 4 &   4.13 & 10.0 &$a+k$ &   1.3 & -61.1 &  -3.0 &  -25.5  &  1.15  & 	0.22  &	 0.787 &   -18.64 \\
2969 &\nodata& 24.94 & 23.52 & 22.59 & 9.0000 &  0 &  \nodata & \nodata & \nodata &  \nodata &  \nodata &  \nodata &   \nodata  & \nodata  & \nodata  &\nodata &    \nodata \\
3072 & 22.98 & 22.42 & 22.30 & 21.46 & 0.5244 &	 1 &   5.26 &  5.0 &$k$   &   1.6 &  \nodata &   0.9 &   -3.7  & \nodata  & \nodata  &	 0.260 &   -19.89 \\
3086 & 23.61 & 22.19 & 21.15 & 19.73 & 0.6428 &	 4 &  12.32 &  0.0 &$k$   &  \nodata &  -1.9 &   3.1 &   -1.8  & \nodata  & 	0.84  &	14.790 &   -22.63 \\
3175 & 26.92 & 24.87 & 24.82 & 22.89 & 9.0000 &  0 &  \nodata & \nodata & \nodata &  \nodata &  \nodata &  \nodata &   \nodata  & \nodata  & \nodata  &\nodata &    \nodata \\
3191 & 24.97 & 25.16 &\nodata&\nodata& 9.0000 &	 0 &  \nodata & \nodata & \nodata &  \nodata &  \nodata &  \nodata &   \nodata  & \nodata  & \nodata  &	\nodata &   \nodata \\
3309 & 25.05 & 24.91 & 24.11 & 23.32 & 9.0000 &	 0 &   6.50 & \nodata & \nodata &  \nodata &  \nodata &  \nodata &   \nodata  & \nodata  & \nodata  &	\nodata &   -17.95 \\
3514 & 25.92 & 24.39 & 23.58 & 22.23 & 0.6438 &	 3 &  11.65 &  0.8 &$k+a$ &   5.0 &   1.1 &  -0.5 &   -3.3  &  1.67  & 	0.94  &	 0.000 &   -19.40 \\
3539 & 24.06 & 23.64 & 23.46 & 23.08 & 9.0000 &	 0 &   3.47 & \nodata & \nodata &  \nodata &  \nodata &  \nodata &   \nodata  & \nodata  & \nodata  &	\nodata &   -13.49 \\
3641 &\nodata&\nodata&\nodata&\nodata& 9.0000 &	 0 &  \nodata & \nodata & \nodata &  \nodata &  \nodata &  \nodata &   \nodata  & \nodata  & \nodata  &	\nodata &   \nodata \\
3650 & 24.49 & 23.97 & 24.18 & 23.81 & 9.0000 &	 0 &   1.90 & \nodata & \nodata &  \nodata &  \nodata &  \nodata &   \nodata  & \nodata  & \nodata  &	\nodata &   -14.02 \\
3705* & 24.49 & 23.84 & 23.85 & 22.55 & 0.7524 &	 4 &   7.03 &  1.5 &$a$   &   2.2 & -67.1 &  -7.5 &  -13.1  &  1.54  & 	0.22  &	 1.925 &   -19.73 \\
3711 & 24.73 & 23.30 & 22.46 & 20.62 & 0.0000 &	 4 &  15.09 & \nodata & \nodata &   0.1 &  \nodata &  \nodata &   \nodata  & \nodata  & \nodata  &	\nodata &   \nodata \\
3833 & \nodata&\nodata&\nodata&\nodata& 9.0000 &	 0 &  \nodata & \nodata & \nodata &  \nodata &  \nodata &  \nodata &   \nodata  & \nodata  & \nodata  &	\nodata &   \nodata \\
3865* & 26.17 & 24.90 & 23.56 & 22.01 & 0.7522 &	 4 &  13.43 &  0.6 &$a+k$ &   5.0 & -11.7 &   9.4 &  -15.6  &  1.57  & 	0.55  &	 0.383 &   -20.18 \\
3922 & 24.49 & 24.25 & 24.07 & 22.81 & 0.9823 &	 2 &   6.19 &  1.0 & \nodata &   2.4 &  \nodata &  \nodata &   \nodata  & \nodata  & \nodata  &	\nodata &   -20.54 \\
4013 & 27.37 & 24.92 & 25.87 & \nodata& 9.0000&	 0 &  \nodata & \nodata & \nodata &  \nodata &  \nodata &  \nodata &   \nodata  & \nodata  & \nodata  &	\nodata &   \nodata \\
4033 & 24.67 & 24.44 & 23.89 & 22.89 & 9.0000 &	 0 &   6.70 & \nodata & \nodata &  \nodata &  \nodata &  \nodata &   \nodata  & \nodata  & \nodata  &	\nodata &   -18.79 \\
4144 & 24.71 & 24.16 & 23.78 & 22.86 & 0.8104 &	 4 &   6.68 &  2.0 & \nodata &   2.2 & -86.8 & -12.6 &   \nodata  &  0.91  & 	0.08  &	 2.815 &   -19.89 \\
4162 & 25.86 & 23.92 & 22.83 & 21.40 & 0.7319 &	 3 &  12.40 &  0.6 &$a+k$ &   5.0 &  -7.1 &  \nodata &   \nodata  &  1.15  & 	0.41  &	 0.392 &   -20.74 \\
4239 &\nodata&\nodata&\nodata&\nodata& 9.0000 &	 0 &  \nodata & \nodata & \nodata &  \nodata &  \nodata &  \nodata &   \nodata  & \nodata  & \nodata  &	\nodata &   \nodata \\
4278* & 24.81 & 24.66 & 24.39 & 23.66 & 0.7687 &	 1 &   3.74 & 10.0 & \nodata &   1.3 & -13.4 &  \nodata &   \nodata  &  0.06  & -0.03  &	 0.206 &   -18.87 \\
4324 &\nodata&\nodata&\nodata&\nodata& 9.0000 &	 0 &  \nodata & \nodata & \nodata &  \nodata &  \nodata &  \nodata &   \nodata  & \nodata  & \nodata  &	\nodata &   \nodata \\
4348 & 23.52 & 23.24 & 23.02 & 22.15 & 0.9809 &	 4 &   4.82 &  1.3 &$a$   &   1.9 & -23.6 &  \nodata &   \nodata  &  0.95  & 	0.56  &	 2.943 &   -21.27 \\
4485 & 25.09 & 24.52 & 22.97 & 22.48 & 0.7035 &	 4 &   7.17 &  2.0 &$k+a$ &   2.4 & -43.7 &   1.0 &    1.0  &  1.16  & 	3.00  &	 1.712 &   -20.07 \\
4524 & 25.42 & 24.27 & 24.19 & 23.62 & 9.0000 &	 0 &  \nodata & \nodata & \nodata &  \nodata &  \nodata &  \nodata &   \nodata  & \nodata  & \nodata  &	\nodata &   \nodata \\
4632 & 22.63 & 22.31 & 22.24 & 22.04 & 9.0000 &	 0 &   2.19 & \nodata & \nodata &  \nodata &  \nodata &  \nodata &   \nodata  & \nodata  & \nodata  &	\nodata &   \nodata \\
\enddata
\end{deluxetable}

\begin{deluxetable}{lrrrrrrrrcrrrrrrrr}
\tablewidth{0pt}
\tabletypesize{\scriptsize}
\tablenum{1b}
\tablecaption{Original Confirmed Cluster Members of Cl 1324+3011}
\tablehead{
\colhead{ID\#} &
\colhead{$AB_B$} &
\colhead{$AB_V$} &
\colhead{$AB_R$} &
\colhead{$AB_I$} &
\colhead{z} &
\colhead{Q} &
\colhead{b} &
\colhead{tau} &
\colhead{Sp} &
\colhead{Age $\tau0.6$} &
\colhead{[OII]} &
\colhead{H$\beta$} &
\colhead{[OIII]} &
\colhead{$J_l$} &
\colhead{$J_u$} &
\colhead{SFR} &
\colhead{$M_{ABB}$}\\
\colhead{(1)} &
\colhead{(2)} &
\colhead{(3)} &
\colhead{(4)} &
\colhead{(5)} &
\colhead{(6)} &
\colhead{(7)} &
\colhead{(8)} &
\colhead{(9)} &
\colhead{(10)} &
\colhead{(11)} &
\colhead{(12)} &
\colhead{(13)} &
\colhead{(14)} &
\colhead{(15)} &
\colhead{(16)} &
\colhead{(17)} &
\colhead{(18)}}
\startdata
 814 & 25.36 & 25.25 & 24.49 & 23.39 & 0.7511 & 3 &  7.58 &  1.6 & \nodata &  2.4 &  -59.4 & \nodata & \nodata & \nodata & \nodata &  0.80 & -18.93\nl
 849 & 24.90 & 24.82 & 24.19 & 23.38 & 0.7506 & 2 &  5.67 &  2.0 &  a  &  1.7 & -111.3 & \nodata & \nodata &  0.12 &  0.07 &  1.86 & -19.06\nl
1680 & 24.99 & 23.39 & 22.35 & 21.02 & 0.7592 & 4 & 11.63 &  0.9 &  k  &  4.0 &    3.4 &    5.0 &   -9.2 &  0.39 &  0.45 &  0.00 & -21.31\nl
1733 & 23.73 & 22.75 & 22.27 & 21.31 & 0.7534 & 4 &  8.14 &  1.6 & k+a &  2.4 &  -13.2 &   -5.9 &   -3.6 &  0.62 &  0.20 &  1.31 & -21.11\nl
1767 & 24.58 & 23.88 & 23.30 & 22.34 & 0.7527 & 4 &  8.04 &  1.4 & a+k &  2.6 &  -37.9 &  -11.2 &  -26.5 &  0.28 &  0.09 &  1.47 & -20.12\nl
1990 & 26.91 & 24.58 & 23.45 & 21.87 & 0.7576 & 4 & 13.57 &  0.4 &  k  &  5.0 &    2.1 &   -1.5 &   -3.4 &  0.46 &  0.44 &  0.00 & -20.37\nl
2114 & 22.92 & 22.57 & 22.13 & 21.37 & 0.7627 & 4 &  5.62 &  2.0 & a+k &  1.7 &   -8.7 &    2.7 &   -3.5 &  0.34 &  0.15 &  1.02 & -21.18\nl
2151 & 23.68 & 22.27 & 21.22 & 19.82 & 0.7528 & 4 & 12.06 &  0.9 &  k  &  4.0 &   -4.9 &    6.3 &    1.5 &  0.52 &  0.53 &  1.35 & -22.43\nl
2186 & 23.13 & 22.88 & 22.60 & 21.89 & 0.7470 & 4 &  4.30 &  5.0 & a+k &  1.4 &   -7.4 &    2.7 &    6.3 &  0.51 &  0.18 &  0.53 & -20.59\nl
2253 & 22.94 & 22.73 & 22.30 & 21.74 & 0.7490 & 4 &  4.37 &  5.0 &  a  &  1.4 &  -26.5 &   -9.1 &   -9.4 &  0.47 &  0.11 &  2.42 & -20.84\nl
2279 & 24.61 & 23.18 & 22.15 & 20.68 & 0.7524 & 4 & 12.52 &  0.7 &  k  &  5.0 &    1.4 &    7.0 &    6.8 &  0.43 &  0.48 &  0.00 & -21.57\nl
2310 & 24.95 & 23.94 & 23.03 & 21.79 & 0.7646 & 4 & 10.77 &  1.1 & a+k &  3.5 &   -2.3 &    7.6 &  -10.8 &  0.70 &  0.32 &  0.12 & -20.59\nl
2331 & 23.12 & 22.98 & 22.80 & 22.20 & 0.7567 & 1 &  2.91 & long &  a  &  0.9 &  -41.2 & \nodata & \nodata &  0.28 &  0.14 &  2.56 & -20.31\nl
2419 & 25.15 & 23.50 & 22.45 & 20.96 & 0.7566 & 4 & 12.82 &  0.7 &  k  &  5.0 &    3.8 &  -12.9 &  -13.4 &  0.49 &  0.57 &  0.00 & -21.30\nl
2452 & 24.25 & 22.49 & 21.46 & 19.96 & 0.7580 & 4 & 12.66 &  0.7 &  k  &  5.0 &   -0.9 &    2.3 &    5.8 &  0.35 &  0.44 &  0.20 & -22.30\nl
2453 & 25.79 & 23.66 & 22.65 & 21.15 & 0.7522 & 4 & 12.71 &  0.7 &  k  &  5.0 &    0.2 &    5.0 &    1.3 &  0.33 &  0.45 &  0.00 & -21.07\nl
2527 & 24.87 & 23.14 & 21.87 & 20.23 & 0.7552 & 4 & 14.03 &  0.2 &  k  &  8.0 &   -2.4 &    7.0 &   -0.2 &  0.38 &  0.47 &  0.38 & -21.97\nl
2611 & 24.47 & 24.30 & 24.07 & 22.39 & 0.7628 & 4 &  8.29 &  1.4 &  k  &  2.4 &   -9.2 &   17.1 &   19.6 &  0.53 &  0.37 &  0.27 & -19.79\nl
2657 & 23.28 & 22.77 & 22.42 & 21.60 & 0.7663 & 4 &  6.07 &  2.0 & k+a &  1.9 &  -28.1 &   -6.7 &  -10.9 &  0.59 &  0.23 &  2.60 & -20.95\nl
2684 & 23.43 & 23.01 & 22.72 & 21.85 & 0.7654 & 1 &  5.62 &  2.0 &  a  &  1.7 &  -39.9 &  -14.6 &  -23.5 &  0.37 &  0.10 &  2.86 & -20.65\nl
2747 & 26.65 & 23.65 & 22.83 & 21.28 & 0.7486 & 4 & 13.00 &  0.4 &  k  &  6.0 &   -0.8 &    0.9 &    9.4 &  0.61 &  0.51 &  0.05 & -20.92\nl
2921 & 25.27 & 23.72 & 22.74 & 21.39 & 0.7564 & 4 & 11.66 &  0.9 &  k  &  4.0 &    1.9 &   11.7 &   -7.3 &  0.45 &  0.48 &  0.00 & -20.91\nl
2932 & 26.38 & 24.66 & 23.76 & 22.80 & 0.7552 & 4 &  9.09 &  1.4 & k+a &  3.0 &   -3.5 &    2.7 &  -13.7 &  0.34 &  0.32 &  0.09 & -19.68\nl
3032 & 25.64 & 23.99 & 23.42 & 21.75 & 0.7618 & 4 & 13.30 &  0.7 & k+a &  5.0 &    2.6 &   10.5 &    1.7 &  0.57 &  0.37 &  0.00 & -20.47\nl
3505 & 23.48 & 22.63 & 21.94 & 20.75 & 0.7494 & 4 &  9.84 &  1.2 & k+a &  3.2 &  -18.2 &   -4.7 &    0.5 &  0.50 &  0.30 &  2.50 & -21.58\nl
3632 & 27.61 & 25.78 & 24.96 & 22.87 & 0.7576 & 0 & 16.80 &  2.0 & \nodata &  4.0 &  -34.8 & \nodata & \nodata & \nodata & \nodata &  0.41 & -19.20\nl
3651 & 23.33 & 22.26 & 21.40 & 20.00 & 0.7513 & 4 & 11.68 &  0.9 & k+a &  4.0 &   -2.5 &    5.7 &   -1.7 &  0.66 &  0.34 &  0.59 & -22.25\nl
3822 & 24.41 & 23.18 & 22.30 & 21.00 & 0.7576 & 4 & 11.20 &  0.9 &  k  &  3.8 &   -0.7 &    2.8 &   -1.6 &  0.50 &  0.35 &  0.07 & -21.33\nl
4061 & 25.21 & 23.24 & 22.42 & 21.07 & 0.7556 & 4 & 11.51 &  0.9 &  k  &  4.0 &   -0.4 &    1.0 &   -0.2 &  0.57 &  0.49 &  0.00 & -21.23\nl
4097 & 24.86 & 23.83 & 23.28 & 22.13 & 0.7524 & 3 &  9.44 &  1.2 & a+k &  3.0 &  -22.3 &   12.0 &    0.5 &  0.98 &  0.48 &  0.91 & -20.21\nl
4439 & 22.12 & 21.77 & 23.40 & 22.81 & 0.7705 & 4 & -1.40 & long & k+a &  0.1 &  -58.5 &   -2.1 &  -62.8 &  0.89 &  0.31 &  6.44 & -20.63\nl
4469 & 21.70 & 21.73 & 21.65 & 21.31 & 0.7706 & 4 &  0.99 & long &  a  &  0.3 &  -60.7 &  -21.9 &  -42.4 &  0.44 &  0.13 & 11.02 & -21.30\nl
\enddata
\tablecomments{These data are taken from Table 2 of Paper IV.}
\end{deluxetable}

\begin{deluxetable}{cccrccrc} 
\tablewidth{0pt}
\tablenum{2}
\tablecaption{Cluster Dynamical Parameters}
\tablehead{
\colhead{} &
\colhead{} &
\colhead{} &
\colhead{} &
\colhead{$M_{PW}$} & 
\colhead{$M_{PM}$} & 
\colhead{$M_{RW}$} & 
\colhead{Radius} \\ 
\colhead{Cluster} & 
\colhead{N$_z$} & 
\colhead{$\overline z$} & 
\colhead{$\sigma$} & 
\multicolumn{3}{c}{($10^{14}~h_{65}^{-1}$ M$_{\odot})$} & 
\colhead{($h_{65}^{-1}$ kpc)}}
\startdata
Cl 1324+3011 & 20 & 0.7567 &1086$^{+231}_{-143}$ & 6.98$^{+2.98}_{-1.85}$ & 8.23$\pm0.51$ & 11.51$\pm0.71$ & 385 \\ 
Cl 1324+3011 & 33 & 0.7561 & 924$^{+142}_{-99} $ & 8.83$^{+2.72}_{-1.89}$ & 8.90$\pm0.37$ & 13.83$\pm0.57$ & 770 \\ 
Cl 1324+3011 & 47 & 0.7564 &1016$^{+126}_{-93} $ & 15.8$^{+4.09}_{-2.91}$ & 29.2$\pm1.46$ & 28.0$\pm1.40$ & Unlimited\\ 
\enddata
\end{deluxetable} 

\clearpage

\voffset +0.80in
\pagestyle{empty}
\begin{deluxetable}{lcrrccclcll}
\scriptsize
\rotate
\tablewidth{0pt}
\tabletypesize{\scriptsize}
\tablewidth{0pt}
\tablenum{3}
\tablecaption{Visual Classifications of Confirmed Cluster Members}
\tablehead{
\colhead{HST ID \#\tablenotemark{a}} &
\colhead{Keck ID \#\tablenotemark{b}} &
\colhead{X\tablenotemark{c}} &
\colhead{Y\tablenotemark{c}} &
\colhead{$m_{best}$\tablenotemark{d}} &
\colhead{Sp\tablenotemark{e}}     &
\colhead{SFRN\tablenotemark{f}}     &
\colhead{Class\tablenotemark{g}}     &
\colhead{D\tablenotemark{h}}         &
\colhead{Interp\tablenotemark{i}}    &
\colhead{Comments\tablenotemark{j}}}
\startdata
     5 & 2151 & 1048.56 &   398.43 &  19.84 & $k$ & 0.144 &	SBb  &   1  &     --	 & \\
     9 & 2518 &  653.98 &   929.79 &  20.22 & $k$ & 0.000 &	E    &   1  &     --	 & \\
    11 & 2452 &  712.57 &   922.66 &  20.44 & $k$ & 0.024 &	E    &   0  &     --	 & \\
    12 & 2279 &  871.28 &   922.87 &  20.48 & $k$ & 0.000 &	S0   &   0  &     I      &  tidal connection w/ \#33\\
    18 & 1680 & 1405.52 &   942.57 &  20.85 & $k$ & 0.000 &	E/S0 &   0  &     --     &  faint comp @2\\
    20 & 2114 & 1002.45 &  1058.73 &  20.96 & $a+k$ & 0.344 &	Sc   &   3  &     --     &  edge-on; bright knot; compact comp @9(\#250)\\
    21 & 2527 &  647.79 &   882.91 &  20.98 & $k$ & 0.061 &	E    &   0  &     I      &  tidal connection to compact comp @1(\#47)\\
    23 & 1733 & 1294.20 &  1556.28 &  21.00 & $k+a$ & 0.470 &	Sc   &	 3  &	  T	 &  tidal arm; several small, faint comps\\
    24 & 2079 & 1073.96 &   729.78 &  21.13 & $k$ & 0.225 &	Sa   &   1  &     --     &  possible faint comp @9\\
    26 & 2453 &  691.61 &  1071.56 &  21.18 & $k$ & 0.000 &	S0   &   1  &     I?     &  slightly asymm disk; close, compact comp @2; 2 other comps @7,11(\#88,113)\\
    29 & 2419 &  742.99 &   974.83 &  21.31 & $k$ & 0.000 &	E    &   0  &     --     &  faint comp @5(\#234); compact comp @7(\#96)\\
    30 & 2170 &  890.38 &  1516.37 &  21.31 & $k$ & 0.000 &	E    &   0  &     --	 & \\
    32 & 2657 &  530.93 &   849.10 &  21.42 & $k+a$ & 1.086 &	P    &   4  &     -- 	 &  ring structure; close comp @4(\#87)\\
    33 & 2253 &  894.87 &   905.35 &  21.49 & $a$ & 1.119 &	Sa   &   3  &     I      &  one arm; asymm disk; tidal connection w/ \#12\\
    40 & 3032 &  259.13 &   252.55 &  21.69 & $k+a$ & 0.000 &	E    &   0  &     --	 & \\
    42 & 2186 &  947.39 &   909.33 &  21.72 & $a+k$ & 0.311 &	P    &   2  &     M,I,T? &  structure in bulge - possible merger; faint tail or galaxy @3; comps @4,8(\#54,617)\\
    51 & 2094 & 1078.96 &   589.21 &  22.02 & $a$ & 0.817 &	P    &   2  &     --     &  structure in nucleus; asymm disk\\
    59 & 1990 & 1128.08 &   965.50 &  22.12 & $k$ & 0.000 &	E    &   1  &     --     &  comps @8,12(\#159,298)\\
    60 & 2310 &  837.45 &   956.88 &  22.12 & $a+k$ & 0.071 &	Sa   &   0  &     --     &  compact comp @11(\#190)\\
    62 & 1767 & 1273.18 &  1479.79 &  22.13 & $a+k$ & 1.315 &	P    &   3  &     M      &  double nucleus in faint disk\\
    69 & 2611 &  539.87 &  1085.77 &  22.26 & $k$ & 0.329 &	S0   &   1  &     --     &  faint, asymm disk\\
   101 & 1934 & 1151.76 &  1190.83 &  22.75 & $a+k$ & 0.732 &	E    &   0  &     --     &  comp @5(\#77)\\
\enddata

\tablenotetext{a}{HST identification number as given in Paper V.}

\tablenotetext{b}{Keck identification number as given here and in Paper IV.}

\tablenotetext{c}{The $x$ and $y$ positions in the final mosaiced HST image
of the PC and the three WFC chips.}

\tablenotetext{d}{Optimal measure of the total magnitude in the F814W
band (see Table 1c of Paper V).}

\tablenotetext{e}{Spectral classification as defined in \S5.2 of Paper IV.}

\tablenotetext{f}{The luminosity-normalized star formation rate in units of 
$M_{\odot}$ yr$^{-1}$ per unit AB$_B$ luminosity (see \S7.2 of Paper IV).}

\tablenotetext{g}{The standard Hubble classification scheme ({\it
e.g.}, E, S0, Sa, Sab etc.).}

\tablenotetext{h}{Disturbance index : 0, normal; 1, moderate
asymmetry; 2, strong asymmetry; 3, moderate distortion; 4, strong
distortion.}

\tablenotetext{i}{Interpretation of disturbance index : M, merger; I,
tidal interaction with neighbor; T, tidal feature; C, chaotic.}

\tablenotetext{j}{Description of galaxy morphology. Here, 
comp is short for  ``companion''; @ indicates position relative to the 
galaxy going clockwise, e.g.\ @7 = ``at 7 o'clock.''}

\end{deluxetable}

\clearpage

\voffset 0in
\hoffset 0in

\end{document}